\documentclass[aps, prb, superscriptaddress, preprint]{revtex4-1}

\usepackage{amsmath}
\usepackage{amsfonts}
\usepackage{amssymb}
\usepackage{graphicx}
\usepackage[final]{pdfpages}
\usepackage{lastpage}
\begin{document}

\title{Ultrafast Electron-Lattice Coupling Dynamics in VO$_2$ and V$_2$O$_3$ Thin Films}
\date{\today}\received{}

\author{Elsa Abreu}
\email{elsabreu@phys.ethz.ch. Previously at Department of Physics, Boston University, Boston MA 02215, USA}
\affiliation{Both authors contributed equally to this work.}
\affiliation{Institute for Quantum Electronics, Department of Physics, ETH Zurich, 8093 Zurich, Switzerland}

\author{Stephanie N. Gilbert Corder}
\affiliation{Both authors contributed equally to this work.}
\affiliation{Department of Physics and Astronomy, Stony Brook University, New York, 11794}

\author{Sun Jin Yun}
\affiliation{Metal-Insulator Transition Lab, ETRI, Daejeon 305-350, South Korea}
\affiliation{School of Advanced Device Technology, University of Science and Technology, Daejeon 305-333, South Korea}

\author{Siming Wang}
\affiliation{Department of Physics, The University of California at San Diego, La Jolla, California 92093, USA}
\affiliation{Center for Advanced Nanoscience, The University of California at San Diego, La Jolla, California 92093, USA}
\affiliation{Materials Science and Engineering Program, The University of California at San Diego, La Jolla, California 92093, USA}
\affiliation{Materials Sciences Division, Lawrence Berkeley National Laboratory, Berkeley, California 94720, USA}

\author{Juan Gabriel Ram\'irez}
\affiliation{Department of Physics, Universidad de los Andes, Bogot\'a 111711, Colombia}

\author{Kevin West}
\affiliation{Department of Materials Science and Engineering, University of Virginia, Charlottesville VA 22904, USA}
\affiliation{Department of Physics, The University of California at San Diego, La Jolla, California 92093, USA}
\affiliation{Center for Advanced Nanoscience, The University of California at San Diego, La Jolla, California 92093, USA}

\author{Jingdi Zhang}
\email{Previously at Department of Physics, Boston University, Boston MA 02215, USA}
\affiliation{Department of Physics, The University of California at San Diego, La Jolla, California 92093, USA}

\author{Salinporn Kittiwatanakul}
\affiliation{Department of Materials Science and Engineering, University of Virginia, Charlottesville VA 22904, USA}

\author{Ivan K. Schuller}
\affiliation{Department of Physics, The University of California at San Diego, La Jolla, California 92093, USA}
\affiliation{Center for Advanced Nanoscience, The University of California at San Diego, La Jolla, California 92093, USA}
\affiliation{Materials Science and Engineering Program, The University of California at San Diego, La Jolla, California 92093, USA}

\author{Jiwei Lu}
\affiliation{Department of Materials Science and Engineering, University of Virginia, Charlottesville VA 22904, USA}

\author{Stuart A. Wolf}
\affiliation{Department of Materials Science and Engineering, University of Virginia, Charlottesville VA 22904, USA}
\affiliation{Department of Physics, University of Virginia,
Charlottesville VA 22904, USA}

\author{Hyun-Tak Kim}
\affiliation{Metal-Insulator Transition Lab, ETRI, Daejeon 305-350, South Korea} 
\affiliation{School of Advanced Device Technology, University of Science and Technology, Daejeon 305-333, South Korea}

\author{Mengkun Liu}
\email{mengkun.liu@stonybrook.edu. Previously at Department of Physics, Boston University, Boston MA 02215, USA}
\affiliation{Department of Physics and Astronomy, Stony Brook University, New York, 11794}

\author{Richard D. Averitt}
\email{raveritt@physics.ucsd.edu. Previously at Department of Physics, Boston University, Boston MA 02215, USA}
\affiliation{Department of Physics, The University of California at San Diego, La Jolla, California 92093, USA}

\begin{abstract}
Ultrafast optical pump - optical probe and optical pump - terahertz probe spectroscopy were performed on vanadium  dioxide (VO$_2$) and vanadium sesquioxide (V$_2$O$_3$) thin films over a wide temperature range.
A comparison of the experimental data from these two different techniques and two different vanadium oxides, in particular a comparison of the electronic oscillations generated by the photoinduced longitudinal acoustic modulation, reveals the strong electron-phonon coupling that exists in the metallic state of both materials.
The low energy Drude response of V$_2$O$_3$ appears more susceptible than VO$_2$ to ultrafast strain control.
Additionally, our results provide a measurement of the temperature dependence of the sound velocity in both systems, revealing a four- to fivefold increase in VO$_2$ and a three- to fivefold increase in V$_2$O$_3$ across the phase transition.
Our data also confirm observations of strong damping and phonon anharmonicity in the metallic phase of VO$_2$, and suggest that a similar phenomenon might be at play in the metallic phase of V$_2$O$_3$.
More generally, our simple table-top approach provides relevant and detailed information about dynamical lattice properties of vanadium oxides, opening the way to similar studies in other complex materials.
\end{abstract}

\maketitle

%%% INTRODUCTION %%%
\section{INTRODUCTION}
\label{intro}

Vanadium oxides are well known examples of materials where the phases are determined by strong interactions between different degrees of freedom.
Such complex systems, where charge, lattice, orbital, and spin contributions can be equally strong and are frequently coupled, exhibit a variety of phenomena including high temperature superconductivity \cite{Hashimoto2014}, colossal magnetoresistance \cite{Jin1994}, multiferroicity \cite{Kubacka2014}, and topological surface states \cite{Wang2013}.
Given the complex nature of these materials, the various phases are generally challenging to investigate experimentally.
Ultrafast time resolved techniques are a successful route to approach these problems \cite{Averitt2002,Orenstein2012}.
In particular, time resolved measurements in different energy ranges have contributed to the understanding of insulator-to-metal transitions (IMTs) in vanadium dioxide (VO$_2$) and vanadium sesquioxide (V$_2$O$_3$) as a function of temperature, pressure, or doping \cite{Misochko1998, Mansart2010, Liu2011, Rodolakis2010, Cavalleri2001, Huber2016, Ocallahan2015, Lantz2017}.
These measurements have also shed light onto the nature of the different insulator phases in both systems \cite{Mansart2010, Pashkin2011, Wegkamp2014, Gray2016}, and the electron-phonon coupling driven acoustic response in V$_2$O$_3$ \cite{Liu2011, Mansart2010}.

Despite many ultrafast and static measurements, the exact mechanisms responsible for the IMTs in vanadium oxides and especially in VO$_2$ remain widely debated \cite{Morin1959, McWhan1969, McWhan1971, Baum2007, Qazilbash2008, Pfalzer2006, Limelette2003, Maurer1999, Baldassarre2008}, namely the contribution of electronic correlations (Mott-Hubbard picture) and of electron-lattice mediated  effects (Peierls model) \cite{Budai2014, Qazilbash2008, Tao2012, Morrison2014, Wegkamp2014}.
It is clear, however, that multiple pathways are possible to initiate pressure and temperature dependent transitions, and that electronic and lattice effects are strongly coupled \cite{Wegkamp2014, Gray2016, Budai2014, Ramirez2015}.
Measurements that dynamically investigate this coupling are therefore essential to assist in understanding the nature of IMTs.

Bulk VO$_2$ undergoes an IMT at 340 K, along with a monoclinic-to-rutile structural transition \cite{Morin1959, Marezio1972} (cf. phase diagram in Fig. S1a \cite{supMaterial}). In the low temperature phase, the vanadium ions dimerize and tilt to form a nonmagnetic insulator \cite{Qazilbash2008}.
Above-bandgap photoexcitation of the monoclinic insulator can promote electrons to anti-bonding states, causing repulsion between the dimerized vanadium ions and lattice expansion, followed by long-range shear rearrangements at the speed of sound \cite{Baum2007, Gray2016}.
Above a critical fluence threshold the metallic phase forms via nucleation and growth \cite{Hilton2007, Lysenko2010, Morrison2014, Wegkamp2014}.
The acoustic and optical phonon landscape changes drastically across the IMT, and phonons have recently been seen to account for 2/3 of the entropy increase at the IMT and to stabilize the metallic phase \cite{Maurer1999, Budai2014}.
These observations suggest that lattice effects play a significant role in driving the IMT in VO$_2$.

V$_2$O$_3$ is a low temperature monoclinic antiferromagnetic insulator which transitions to a paramagnetic rhombohedral metal above 155 K \cite{Morin1959, McWhan1970} (cf. phase diagram in Fig. S1b \cite{supMaterial}).
Initial reports stated that the system exhibits a classical Mott-Hubbard transition where electrons localize to form the insulating phase \cite{McWhan1969,McWhan1971}.
However, more recent work points to electron-lattice effects contributing strongly to the first order IMT in the undoped compound \cite{Budai2014,Pfalzer2006}, and questions the equivalence of temperature and pressure routes in driving the phase transition in doped V$_2$O$_3$ \cite{Rodolakis2010}.
The strong strain-dependence of metallic V$_2$O$_3$ manifests via the large influence of ultrafast acoustic modulations on the spectral weight redistribution dynamics, particularly in the far infrared region of the spectra corresponding to the low energy Drude weight \cite{Liu2011}.

In this work, we present a comparative ultrafast pump-probe study of photoinduced acoustic effects in VO$_2$ and V$_2$O$_3$, present in both optical reflectivity and terahertz (THz) conductivity dynamics.
Transient reflectivity measurements at 1.55 eV are sensitive to interband transitions in VO$_2$ and V$_2$O$_3$, with insulating gaps of 0.6 eV and 0.5 eV, respectively \cite{Qazilbash2008}, while THz probes the quasiparticle dynamics, effectively yielding a dynamical measurement of the dc conductivity.

Our results show that, similar to V$_2$O$_3$ \cite{Liu2011, Mansart2010}, VO$_2$ exhibits a modulation of the electronic response due to photoinduced acoustic effects.
This influence of small structural variations on the electronic behavior attests to significant electron-phonon coupling.
Comparing both materials, we observe that V$_2$O$_3$ is significantly more amenable than VO$_2$ to Drude weight modulation via acoustic wave propagation.
Also, thin film samples with varying defect density exhibit different static and dynamic electronic responses.
Thin films with fewer defects have properties closer to bulk and are therefore characterized by a larger THz conductivity in the metallic state and by larger photoinduced conductivity variations.
In contrast to this electronic behavior, the acoustic signatures we observe appear to be quite robust against varying defect density in the thin films.
Differences in the response to photoinduced acoustic excitation between VO$_2$ and V$_2$O$_3$, and between nominally equivalent samples, could be related to the strong and distinct influence of defects on the electronic response of these materials \cite{Ramirez2015}.

In addition, our data provide a temperature dependent measure of the sound velocity in both VO$_2$ and V$_2$O$_3$, revealing an increase in the sound velocity across the insulator-to-metal transition, four- to fivefold for VO$_2$ and three- to fivefold for V$_2$O$_3$.

We also verify that acoustic damping increases in the metallic phase of VO$_2$, compared to the insulating phase, in agreement with previous reports that present phonon entropy as a stabilization mechanism for the metallic phase \cite{Budai2014}.
V$_2$O$_3$ exhibits a similarly increased damping in the metallic phase, hinting that phonon anharmonicity might also be much stronger than in the insulating phase, and that phonon effects might play a significant role in the IMT of this material as well.

%%% EXPERIMENTAL METHODS %%%
\section{EXPERIMENTAL METHODS}
\label{exp}

Transient optical reflectivity measurements were performed using 35 fs pulses at 800 nm (1.55 eV) from a 1 KHz repetition rate, 3 W average power Ti:sapphire regenerative amplifier.
Pump and probe fluences were set at 1-4 mJ/cm$^2$ and $<10~\mu$J/cm$^2$, respectively, and both beams were set at approximately normal incidence to the sample.
THz pulses 1 ps in duration were generated via optical rectification and detected by electro-optic sampling in ZnTe.
In contrast to the optical measurements at 1.55 eV, the THz pulses (0.1 - 2.5 THz) effectively probe the low energy Drude response \cite{Qazilbash2008}.
Transient optical reflectivity  measurements at 800 nm, yielding $\Delta$R/R, and THz conductivity measurements, yielding $\Delta \sigma$ \cite{Averitt2002}, were performed with the same pump conditions.

Of particular relevance to this work is the ability to generate coherent acoustic phonons using ultrafast optical excitation.
The ultrafast above bandgap optical pulse is absorbed near the surface causing a localized temperature increase.
Thermal expansion leads to a transient stress which launches a strain wave in the material.
The propagating acoustic phonon modifies the refractive index and can therefore be detected optically via modulation of $\Delta$R/R, at 800 nm, and of $\Delta \sigma$, around 1 THz.
When the film thickness is on the order of the longitudinal acoustic phonon wavelength an acoustic standing wave is generated and detected instead \cite{Ge2014}.
Analytical \cite{Thomsen1984, Thomsen1986} and conceptual \cite{Ge2014} aspects of the generation and detection of coherent acoustic modulations can be found in the literature \cite{Thomsen1984, Thomsen1986, Ge2014}.

The samples consist of 75 nm and 50 nm thick films of VO$_2$ on c-cut Al$_2$O$_3$ \cite{West2008}, 75 nm thick films of V$_2$O$_3$ on c-cut Al$_2$O$_3$ \cite{Yun2009}, and 95 nm thick films of V$_2$O$_3$ on r-cut Al$_2$O$_3$ \cite{Ramirez2015}.
The VO$_2$ films have a well defined out of plane rutile c-axis, and three preferred in-plane orientations due to the hexagonal symmetry of the c-cut Al$_2$O$_3$ substrate \cite{Kittiwatanakul2014_2}.
The 75 nm V$_2$O$_3$ film has a well-defined rhombohedral out of plane [110] axis but is polycrystalline in plane, while the 95 nm film is nearly single crystalline throughout.
Details of the fabrication can be found elsewhere \cite{West2008, Yun2009, Ramirez2015}.
Unless otherwise specified, the results presented in this paper correspond to the 75 nm films.

The sample temperature was set using a continuous flow cryostat, pumped down to 10$^{-6}$ mbar and equipped with a heating stage.

%%% RESULTS %%%
\section{RESULTS}
\label{results}

\begin{figure} [htb]
\begin{center}
\includegraphics[width=0.4\textwidth,keepaspectratio=true]{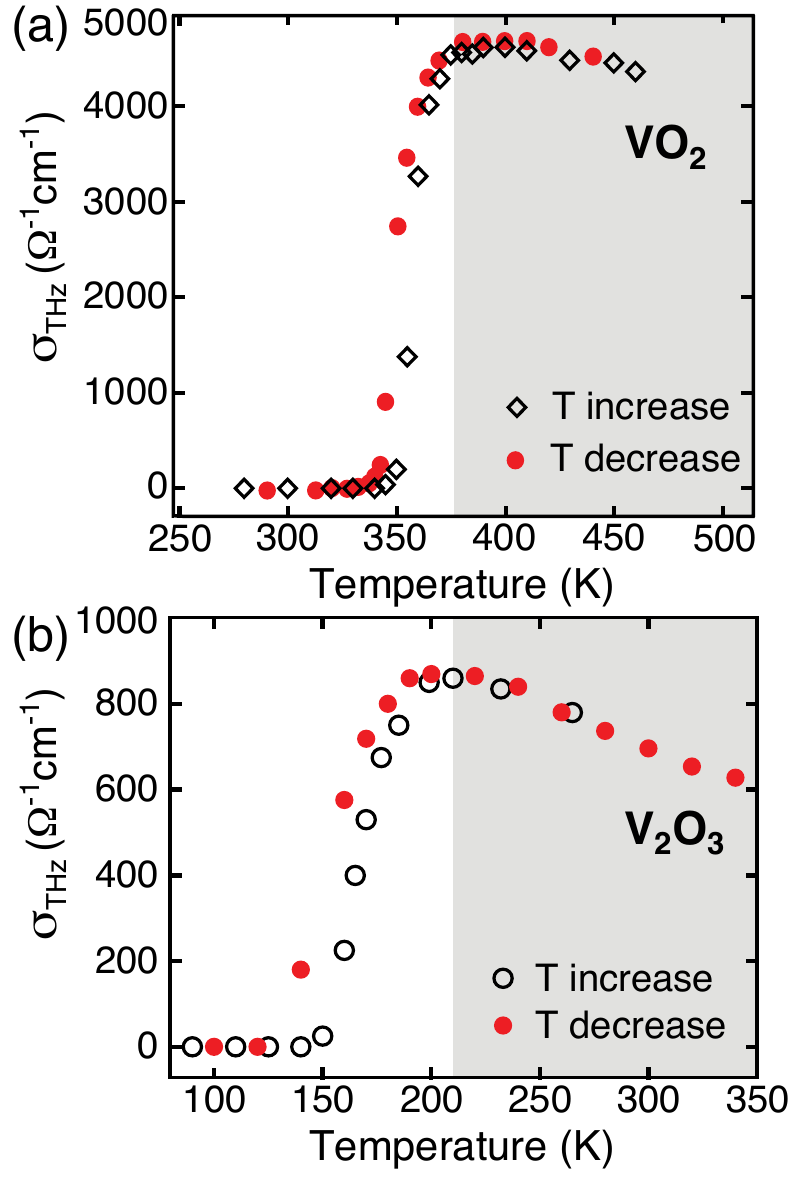}
\caption{
Temperature dependence of the static far-infrared conductivity for the 75 nm thick (a) VO$_{2}$ and (b) V$_2$O$_3$ thin films.
Open black (solid red) symbols indicate increasing (decreasing) temperature.
Both samples display hysteresis, characteristic of first order phase transitions happening at 360 K in VO$_2$ and at 160 K in V$_2$O$_3$.
The fully metallic region is shaded gray.
}
\label{static}
\end{center}
\end{figure}

Figure \ref{static} shows the static THz conductivity of the 75 nm VO$_2$ and V$_2$O$_3$ films measured using THz spectroscopy.
Fig. \ref{static}a shows that the IMT in the VO$_2$ film occurs at $T_{IMT} = 360$ K, and that a maximum conductivity of 4800 ($\Omega$cm)$^{-1}$ is obtained at 380 K.
V$_2$O$_3$ is seen from Fig. \ref{static}b to have $T_{IMT} = 160$ K, with a peak conductivity of 900 ($\Omega$cm)$^{-1}$ at 200 K.
These measurements demonstrate that our VO$_2$ and V$_2$O$_3$ films behave similarly to bulk single crystals, which attests to the generality of our conclusions.
Above 200 K the V$_2$O$_3$ metallic state conductivity decreases dramatically (Fig. S1 \cite{supMaterial}), beyond what would be expected from an increased electron-phonon scattering rate, due to the strongly correlated nature in this ``bad" metal region \cite{Qazilbash2006, Qazilbash2008}.
The relative conductivity decrease in the VO$_2$ metallic state above 380 K is smaller compared to V$_2$O$_3$, with a 6\% conductivity drop per 50 K for VO$_2$ and an 11 \% conductivity drop per 50 K for V$_2$O$_3$.
This is consistent with suggestions that electronic correlations have a smaller impact on the metallic phase of VO$_2$ compared to metallic V$_2$O$_3$ \cite{Basov2011, Stewart2012}, which enters a crossover region to a pure Mott insulator phase above 450 K (Fig. S1 \cite{supMaterial}).
Corresponding temperature-dependent static conductivity data for the 95 nm V$_2$O$_3$ film are shown in Fig. S2 \cite{supMaterial}.

\begin{figure} [htb]
\begin{center}
\includegraphics[width=0.4\textwidth,keepaspectratio=true]{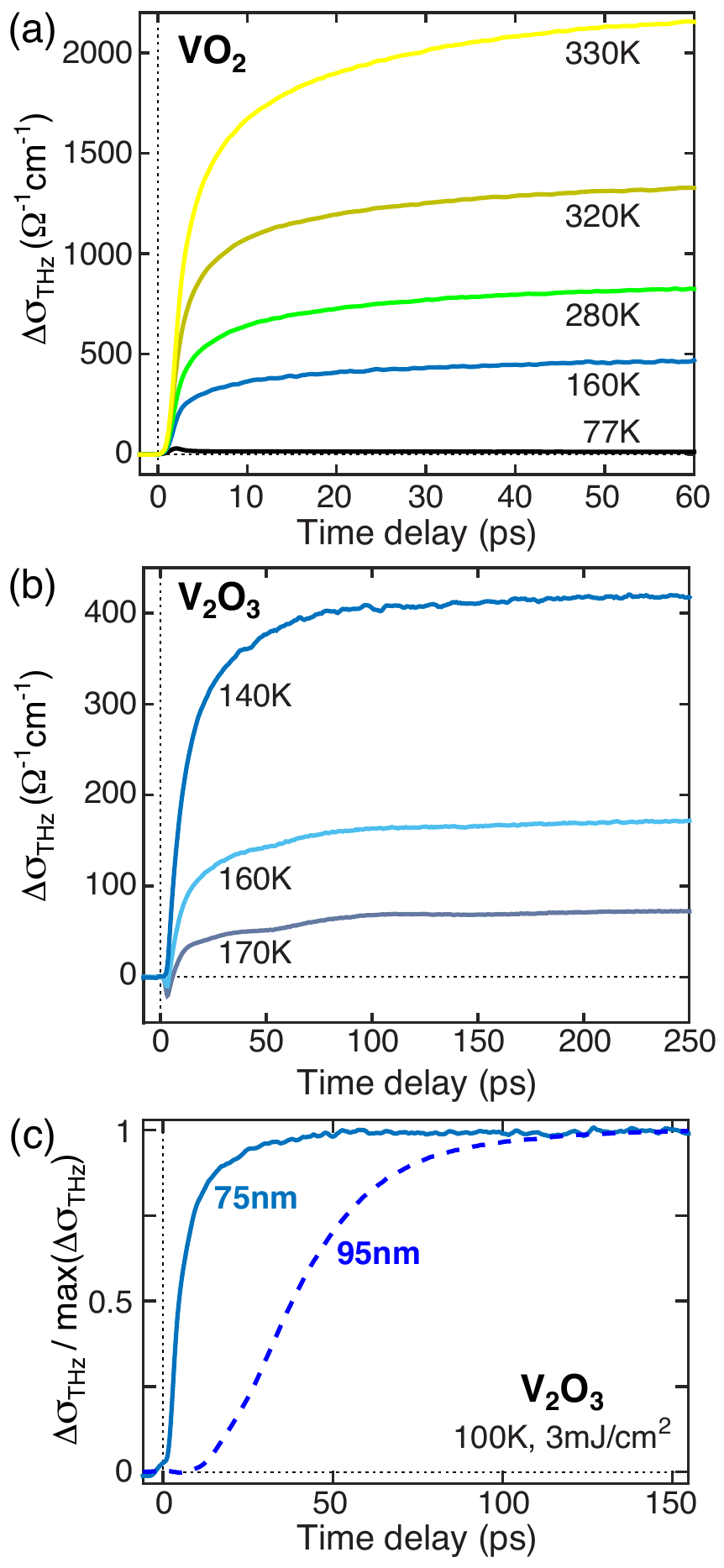}
\caption{
Transient $\Delta\sigma$ of 75 nm thick VO$_2$ (a), 75nm thick V$_2$O$_3$ (b), and 75 nm and 95 nm thick V$_2$O$_3$ (c) for temperatures below $T_{IMT}$ at a pump fluence of 3.8 mJ/cm$^2$ (a), 1 mJ/cm$^2$ (b) and 3 mJ/cm$^2$ (c).
Starting in the insulating phase, the THz response consists in a transient conductivity increase which corresponds to the IMT.
(c) compares the normalized $\Delta\sigma(t)$ for the 95 nm and 75 nm V$_2$O$_3$ films at 100 K, revealing a significant difference in IMT dynamics at short time delays which arises from different defect densities in the films.
}
\label{pplowT}
\end{center}
\end{figure}

\begin{figure} [h!tb]
\begin{center}
\includegraphics[width=0.36\textwidth,keepaspectratio=true]{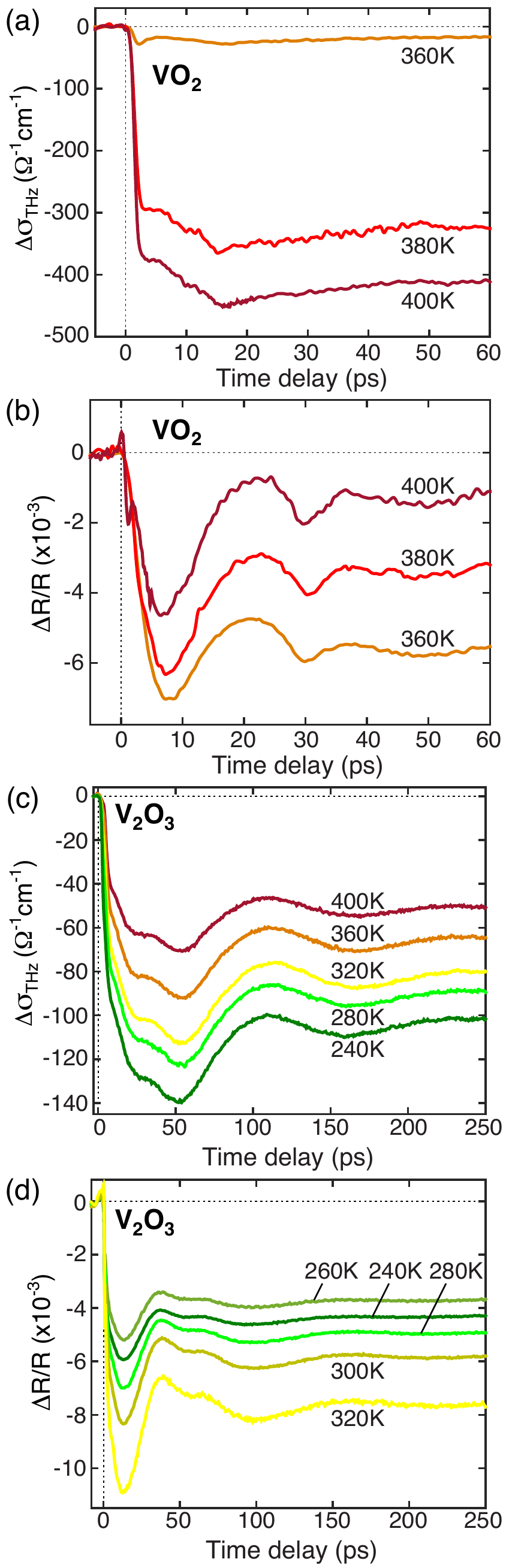}
\caption{
Transient $\Delta\sigma$ (a, c) and $\Delta R/R$ (b, d) of 75 nm thick VO$_2$ (a, b) and V$_2$O$_3$ (c, d) for temperatures above $T_{IMT}$ at a pump fluence of 3.8 mJ/cm$^2$ (a, b) and 1 mJ/cm$^2$ (c, d).
Shown are the metallic reflectivity responses, dominated by acoustic signatures, and the metallic conductivity responses, also with clear acoustic signatures.
}
\label{pphighT}
\end{center}
\end{figure}

Figures \ref{pplowT}a, \ref{pphighT}a and \ref{pphighT}b show the transient conductivity ($\Delta\sigma$) and transient reflectivity at 800 nm ($\Delta R/R$) of VO$_2$ at low (Fig. \ref{pplowT}a) and high (Figs. \ref{pphighT}a and \ref{pphighT}b) initial temperatures for a pump fluence of 3.8 mJ/cm$^{2}$. 
From the conductivity dynamics at $T<T_{IMT}$, shown in Fig. \ref{pplowT}a, it is clear that the maximum value achieved for the transient conductivity $\Delta\sigma(t)$ increases with initial temperature. 
For $T=77$ K, $\Delta\sigma(t)$ recovers in less than 10 ps, whereas for $T>160$ K the deposited energy is sufficient to thermally stabilize the metallic phase beyond our 350 ps measurement window \cite{supMaterial}.
As previously observed, the tens of ps timescale for the conductivity increase is considerably larger than the $<$1 ps electron-phonon thermalization time (Fig. S9 \cite{supMaterial}) due to the nucleation and growth process that accompanies the IMT \cite{Hilton2007, Abreu2015}.

For $T>T_{IMT}$ the metallic phase dominates the transient photothermal response of VO$_2$.
$\Delta\sigma$, shown in Fig. \ref{pphighT}a, decreases following photoexcitation in agreement with Fig. \ref{static}a, and so does $\Delta R/R$ (Fig. \ref{pphighT}b).
Most significantly, the high temperature dynamics of both $\Delta\sigma$ and $\Delta R/R$ exhibit oscillatory components (clearly isolated below, in Figs. \ref{optp}a and \ref{opop}b)  which are direct signatures of acoustic wave propagation, similar to those reported in V$_2$O$_3$ \cite{Mansart2010, Liu2011}.
Budai \textit{et al.} \cite{Budai2014} have observed that strongly anharmonic phonons rather than electronic effects stabilize the metallic phase of VO$_2$.
The results shown here confirm the strength of electron-phonon coupling in metallic VO$_{2}$ since acoustic signatures are seen not only in $\Delta R/R$, as expected, but also in $\Delta\sigma$, a clear indication that lattice dynamics modulate the Drude response of the system above $T_{IMT}$.

As a counterpart to the data on VO$_2$, Figs. \ref{pplowT}b, \ref{pphighT}c and \ref{pphighT}d show $\Delta\sigma$ and $\Delta R/R$ for V$_2$O$_3$ at temperatures below (Fig. \ref{pplowT}b) and above (Figs. \ref{pphighT}c and \ref{pphighT}d) $T_{IMT}$ for a pump fluence of 1 mJ/cm$^{2}$.
V$_2$O$_3$ also exhibits a tens of ps $\Delta\sigma$ increase characteristic of a percolative IMT (Fig. \ref{pplowT}b).
The decrease in $\Delta\sigma$ with increasing initial temperature stems solely from the fact that temperatures closer to $T_{IMT}$ are used compared to VO$_2$, so that the initial state already has a finite conductivity (Fig. \ref{static}) and the $\Delta\sigma$ response saturates as the full metallic state is reached.
Photoexcitation of the system in the metallic phase at $T>T_{IMT}$ produces a decrease of both $\Delta\sigma$ and $\Delta R/R$ in addition to clearly defined temperature dependent acoustic signatures, as previously reported \cite{Liu2011, Mansart2010}.

Figure \ref{pplowT}c compares the normalized $\Delta\sigma(t)$ for the 95 nm and 75 nm V$_2$O$_3$ films at 100 K for a pump fluence of 3 mJ/cm$^{2}$.
The 95 nm film has a metallic conductivity more than twice as large as the 75 nm film (Fig. \ref{static}b and Fig. S2 \cite{supMaterial}), which is likely due to its single crystalline rather than polycrystalline nature and to its consequent smaller defect density.
As discussed in an earlier publication \cite{Abreu2015}, the defect density has a strong influence not only on the static properties but also on the transition dynamics of V$_2$O$_3$ thin films since it affects the nucleation and growth of metallic domains in the insulating phase.
For V$_2$O$_3$ and VO$_2$, it is important to distinguish between ultrafast transition dynamics that occur at the microscopic level, independent of the nucleation and growth process, and those that occur at the mesoscopic level where the response is dominated by nucleation and growth.
Microscopic effects have been extensively discussed, particularly in the case of VO$_2$ \cite{Hilton2008,Qazilbash2008,Liu2013,Donges2016}.
Using a THz probe, which is sensitive to the mesoscale dynamics, we observe fast $\Delta\sigma(t)$ transients that start during photoexcitation by the pump pulse for the 75 nm film, whereas the 95 nm film exhibits slower variations and a delayed onset relative to the pump arrival time (Fig. \ref{pplowT}c).
These significant differences in conductivity dynamics cannot be explained by the minimal difference in film thickness.
Rather, they occur due to the 95 nm film containing fewer defects than the 75 nm  film, and hence fewer defect-induced preferential nucleation sites, which slows down the photoinduced IMT in the thicker sample \cite{Abreu2015}.
Such mesoscopic effects occur independently in addition to any microscopic modifications, and their effect on the dynamics must be taken into account in studies of both V$_2$O$_3$ and VO$_2$.

Despite differences in IMT timescales (Figs. \ref{pplowT}c and S6 \cite{supMaterial}), high temperature dynamic data for the 50 nm VO$_2$ film and the 95 nm V$_2$O$_3$ film are comparable to the 75 nm films of each material and consequently are shown in Figs. S4 and S7 of the Supplemental Material \cite{supMaterial}.

Finally, $\Delta R/R$ dynamics for VO$_2$ and V$_2$O$_3$ at $T<T_{IMT}$ also exhibit acoustic oscillations characteristic of the insulating phase.
The full dynamic insulating responses have many contributing factors and are therefore less straightforward to analyze than the metallic response. 
As the present work focuses on the acoustic component, the full low temperature $\Delta R/R$ signals are shown in Fig. S3 \cite{supMaterial}.
The acoustic signatures of the insulating phase can, however, be compared with the metallic phase results of Figs. \ref{pphighT}b and \ref{pphighT}d, as shown in Fig. \ref{opop} and discussed in Section \ref{analysis} below.

%%% ANALYSIS AND DISCUSSION %%%

\section{ANALYSIS AND DISCUSSION}
\label{analysis}

\begin{figure} [h!tb]
\begin{center}
\includegraphics[width=0.43\textwidth,keepaspectratio=true]{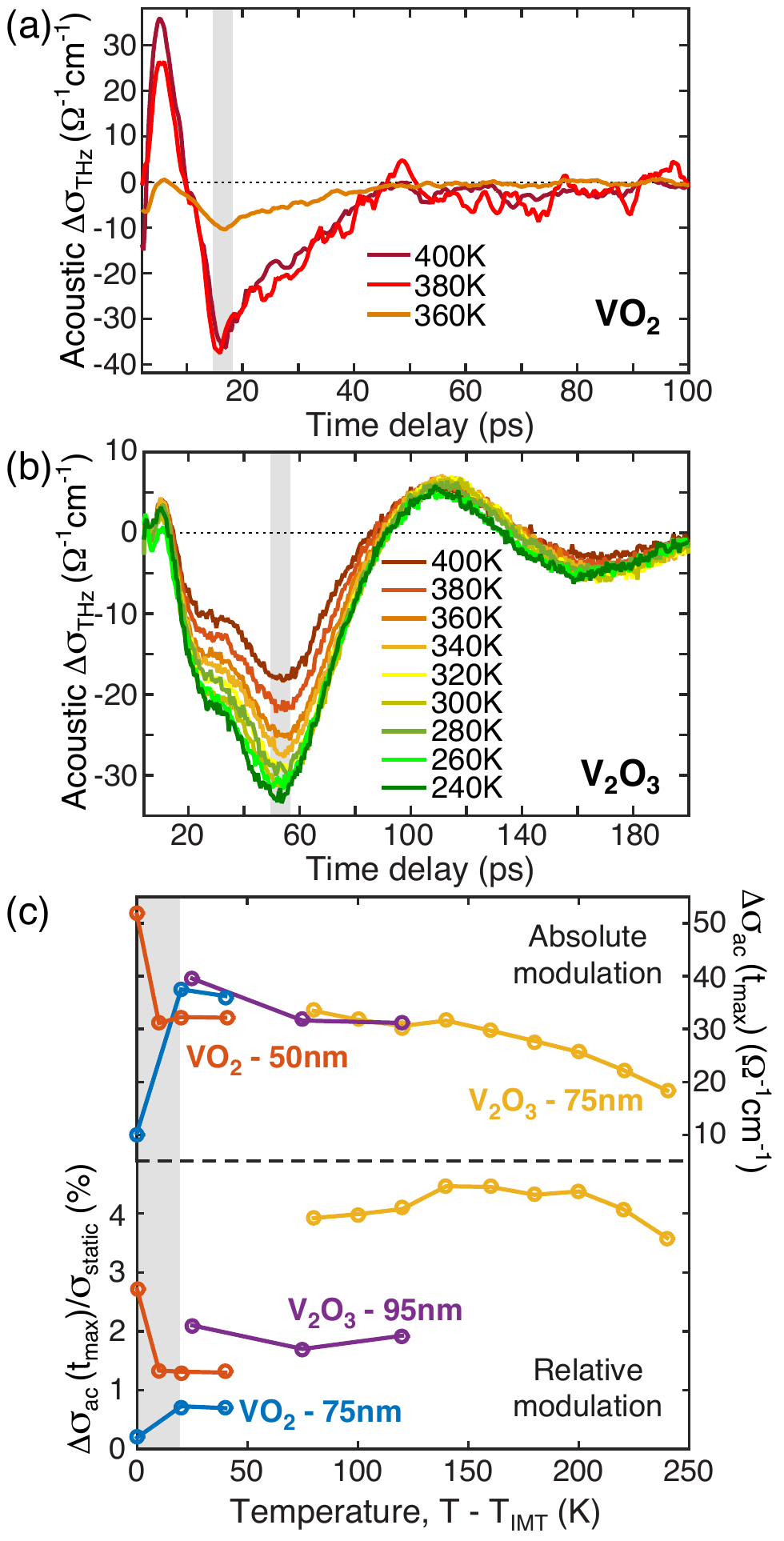}
\caption{
Background subtracted $\Delta \sigma$ acoustic oscillations for (a) VO$_2$ and (b) V$_2$O$_3$, corresponding to the data from Figs. \ref{pphighT}a and \ref{pphighT}c.
$\Delta\sigma(t)$ data for the VO$_2$ film were smoothed using a five point (1 ps) moving average.
Vertical gray bars indicate the time delays at which the values in (c) were calculated \cite{supMaterial}.
(c) Bottom: Temperature dependence of the maximum acoustic modulation of $\Delta \sigma$, normalized by the static conductivity at the corresponding temperature (Figs. \ref{static} and S2 \cite{supMaterial}).
Top: Temperature dependence of the absolute maximum acoustic modulation of $\Delta \sigma$.
Data is shown for all four VO$_2$ and V$_2$O$_3$ films.
A fluence of 3.8 mJ/cm$^{2}$ was used for the 75 nm thick VO$_2$ film (blue), 4 mJ/cm$^{2}$ for the 50 nm VO$_2$ (orange), 1 mJ/cm$^{2}$ for the 75 nm V$_2$O$_3$ (yellow) and 3 mJ/cm$^{2}$ for the 95 nm V$_2$O$_3$ (purple).
The gray shaded region corresponds to temperatures $T<T_{IMT}+20$ K, where the influence of the IMT hysteresis could affect the results.
}
\label{optp}
\end{center}
\end{figure}

\begin{figure} [h!tb]
\begin{center}
\includegraphics[width=1.\textwidth,keepaspectratio=true]{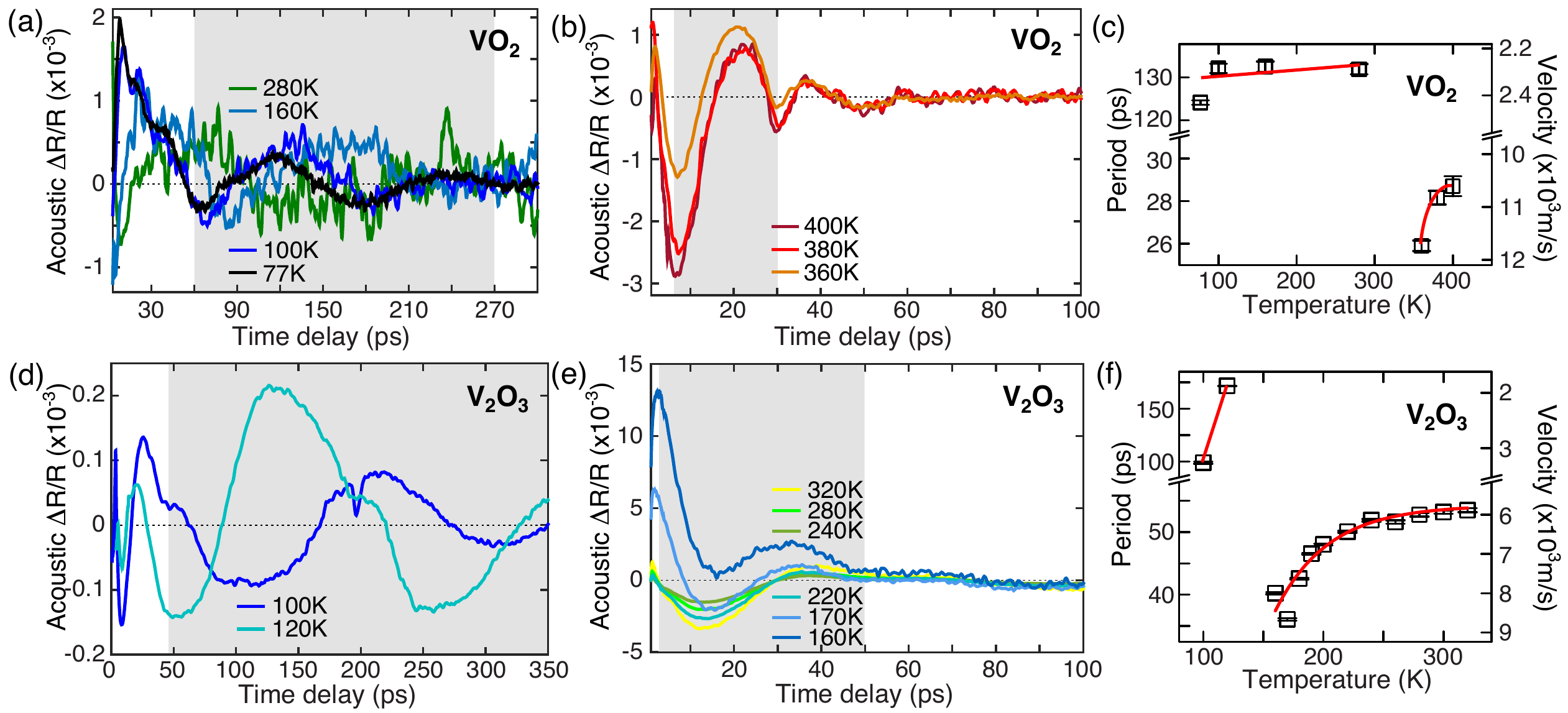}
\caption{
Background subtracted $\Delta R/R$ acoustic oscillations for insulating (a) and metallic (b) VO$_2$ and for insulating (d) and metallic (e) V$_2$O$_3$, corresponding to the data in Figs. \ref{pphighT}b, \ref{pphighT}d and S3 \cite{supMaterial}.
Plots to the right show the temperature dependence of the acoustic oscillation period and of the corresponding sound velocity for VO$_2$ (c) and for V$_2$O$_3$ (f), calculated by fitting the shaded regions of a, b, d and e with a damped sinusoid.
Red lines are linear fits (c, f: insulating phase), an exponential guide to the eye (c: metallic phase) and an exponential fit (f: metallic phase).
}
\label{opop}
\end{center}
\end{figure}

In order to analyze the acoustic dynamics in more detail, we subtract all exponential, non-oscillatory contributions to the metallic data in Fig. \ref{pphighT}.
This procedure is described in the Supplemental Material \cite{supMaterial}, and yields the acoustic contribution to $\Delta\sigma(t)$, shown in Fig. \ref{optp}, and to the $\Delta R/R$ dynamics, shown in Fig. \ref{opop}.
The lack of acoustic signatures in the insulating phase $\Delta\sigma(t)$ response in Figs. \ref{pplowT}a and \ref{pplowT}b is expected since the THz probe is only sensitive to the metallic volume fraction \cite{Abreu2015}.

The acoustic component of the $\Delta\sigma$ dynamics is shown in Fig. \ref{optp}a and \ref{optp}b for VO$_2$ and V$_2$O$_3$, respectively.
The  modulation is significantly longer lived in V$_2$O$_3$, an indication that phonon damping is smaller in V$_2$O$_3$ than in VO$_2$.
The significance of phonon damping for the properties of VO$_2$ and V$_2$O$_3$ will be discussed in more detail below.

In Fig. \ref{optp}c, the maximum $\Delta\sigma$ oscillation amplitude for both VO$_2$ and V$_2$O$_3$ is plotted as a function of $T-T_{IMT}$.
The maximum is estimated at the time delays marked by vertical gray bars in Figs. \ref{optp}a and \ref{optp}b to avoid potential residual contributions from non-acoustic effects at shorter time delays.
This choice of temperature scale enables a direct comparison between the different samples and materials.
The bottom part of Fig. \ref{optp}c shows the oscillation amplitude maxima normalized by the static conductivity at that temperature \cite{supMaterial}, while the top shows the non-normalized values.
For $T>T_{IMT}+20$ K, i.e. well outside the hysteresis region (Fig. \ref{static}), the absolute value of the maximum conductivity modulation (top of Fig. \ref{optp}c) decreases with increasing temperature for all films.
This decrease is consistent with the negative slope of the static conductivity observed with increasing temperature in the metallic phase (Figs. \ref{static} and S2 \cite{supMaterial}).
Interestingly, the conductivity oscillation amplitudes are very similar for all the samples examined, even after accounting for the differences in pump fluence (the fluence dependence for the two V$_2$O$_3$ samples is discussed in Section II G and Fig. S8 of the Supplemental Material \cite{supMaterial}).
In particular, oscillation amplitudes are comparable for VO$_2$ and V$_2$O$_3$, as well as for samples of the same material with different static conductivities. 
The normalized plot in Fig. \ref{optp}c (bottom) enables a more direct comparison of VO$_2$ and V$_2$O$_3$ results.
It is clear that acoustic modulations in the 75 nm thick V$_2$O$_3$ are stronger than in VO$_2$, which points to larger electron-phonon induced strain modulation.
This conclusion is further strengthened if the different pump fluences are taken into account when analyzing the data (cf. Section II G and Fig. S8 of the Supplemental Material \cite{supMaterial}).
The 95 nm thick V$_2$O$_3$ film does not show such a strikingly higher effect compared to VO$_2$.

It is not trivial to relate the acoustic oscillation period in $\Delta\sigma(t)$ with the sound velocity in the system since the THz probe, with its $\sim$300 $\mu$m wavelength, effectively probes an average acoustic modulation of the conductivity in the sub-micron thick films \cite{Liu2011}.
However, an approximate sound velocity value can be directly determined using oscillations of $\Delta R/R$, the transient reflectivity at 800 nm \cite{Thomsen1984}.

We focus on the acoustic oscillations which are isolated in Fig. \ref{opop}, obtained from the data in Figs. \ref{pphighT}b and \ref{pphighT}d for $T>T_{IMT}$, and in Fig. S3 \cite{supMaterial} for $T<T_{IMT}$.
Acoustic modulations of $\Delta R/R$ for $T<T_{IMT}$ are more difficult to observe in V$_2$O$_3$ than in VO$_2$ as the current measurements are limited to temperatures above 80 K.
Indeed, the lower value of $T_{IMT}$ for V$_2$O$_3$ compared to VO$_2$, as well as the fact that the latent heat is smaller and therefore the IMT can be driven at the same fluence for lower temperatures relative to the respective $T_{IMT}$, means that the acoustic response of the insulating phase in V$_2$O$_3$ is quickly masked by phase transition dynamics.
This problem could be circumvented in future work by analyzing the response at lower temperatures or by studying doped V$_2$O$_3$ samples.

The period of the reflectivity oscillations shown in Figs. \ref{opop}c and \ref{opop}f is determined by fitting the shaded regions in the corresponding reflectivity data with a damped sinusoid.
The fidelity of the fit is validated by the near unity adjusted R$^2$ values (cf. Section II D of the Supplemental Material \cite{supMaterial}).
Given that the 75 nm thickness of the films is on the order of acoustic phonon wavelengths \cite{Seikh2006}, the observed oscillations are essentially reflections of the acoustic wave that propagates in the sample between the film surface and the film-substrate interface, as described in Section \ref{intro}.
The corresponding sound velocity can therefore be calculated from the period, $\tau$, using $v_{sound} = 4 d / \tau$ \cite{Thomsen1984, Ge2014}, where $d$ is the film thickness (a more detailed analysis is given in Section III and Fig. S9 of the Supplemental Material, for a thicker film \cite{supMaterial}).

The most striking effect in our data is that the sound velocity associated with these oscillations, apart from a slow and monotonic increase with temperature, exhibits an abrupt increase at $T_{IMT}$ when going from the insulating to the metallic phase, by a factor of 4-5 in VO$_2$ and 3-5 in V$_2$O$_3$.
This is seen clearly in Figs. \ref{opop}c and \ref{opop}f.
Such a large sound velocity variation is not unexpected in vanadium oxides, where lattice constants change by about 1$\%$ across the structural phase transition that accompanies the IMT \cite{Kucharczyk1979, McWhan1970, supMaterial}.
V-O and V-V bond lengths change by up to 4$\%$ across the structural transition, stabilizing an insulating phase below $T_{IMT}$ or a metallic phase above $T_{IMT}$ \cite{Dernier1970, Marezio1972, supMaterial}.

Sound velocity measurements for the metallic phase of VO$_2$ yield approximately 4x$10^3$ m/s, measured along the rutile c-axis by Maurer \emph{et al.} \cite{Maurer1999}, or 8x$10^3$ m/s, determined from the phonon dispersion of the longitudinal acoustic mode along the $\Gamma$-Z direction in Budai \emph{et al.} \cite{Budai2014}.
In the metallic phase of V$_2$O$_3$ Seikh \emph{et al.} \cite{Seikh2006} and Yelon \emph{et al.} \cite{Yelon1979} report 8x$10^3$ m/s, corresponding to longitudinal acoustic mode propagation.
The VO$_2$ and V$_2$O$_3$ metallic sound velocities measured in Fig. \ref{opop} are consistent with the order of magnitude reported previously.
Some discrepancies are expected, as discussed below. 

Sample to sample variation in vanadium oxides, and in particular in vanadium oxide thin films, is well known and must be taken into account when analyzing and comparing different data sets \cite{Liu2015}.
In particular, variation in sound velocity values is a result of the different strain environments and crystallinity of the samples.
The strain sensitivity of  VO$_2$ material properties, in particular, is widely reported \cite{Goodenough1971, Wu2006, Park2013, Kittiwatanakul2014, Liu2015} and is caused by large differences in lattice constant values along different crystallographic axes with large variations of these values across the IMT (Table SI \cite{supMaterial}).
For the measurements presented here, the sound velocity determination could be affected by acoustic waves propagating at an angle to the film normal, by substrate-induced strain in the growth direction, as well as by contributions from other acoustic modes in the system, in particular non-longitudinal waves.

Analyzing Figs. \ref{opop}b and \ref{opop}e from the perspective of acoustic modulation damping reveals that the amplitude of the metallic oscillations decays two times faster in VO$_2$ than in V$_2$O$_3$ (Fig. \ref{opop}), with damping times of about 10 ps and 20 ps, respectively \cite{supMaterial}.
Damping effects are thus stronger in the metallic phase of VO$_2$ compared to V$_2$O$_3$.
The observations from Budai \emph{et. al.} \cite{Budai2014} of increased phonon damping in the metallic phase compared to the insulating phase are verified by our analysis on VO$_2$, where the damping time decreases from 145 ps in the insulator to 10 ps in the metal.
The results for V$_2$O$_3$ show a decrease in damping time from 225 ps in the insulating phase to 20 ps in the metallic phase, suggesting a similar influence of phonon anharmonicity to be at play in the metallic phase of V$_2$O$_3$.
It should be noted that an additional source of damping exists, due to transmission of the acoustic wave into the substrate.
Transmission losses will depend on how well acoustic impedances are matched between the film and the substrate, and can potentially lead to an increase of the effective damping time associated with the VO$_2$ or V$_2$O$_3$ material responses alone.
Quantitative comparisons with damping times obtained from other methods must therefore be done with care.
A detailed analysis of the observed damping is presented in Section II D of the Supplemental Material \cite{supMaterial}.

Figures \ref{optp} and \ref{opop} therefore enable us to conclude that acoustic signatures are qualitatively the same and quantitatively within a factor of two of each other in VO$_2$ and V$_2$O$_3$.
This is so despite differences in the mechanism that drives the IMT, in particular the larger effect expected from electronic correlations in V$_2$O$_3$ compared to VO$_2$, and in the latent heat associated with the IMT (65 J/cm$^3$ for V$_2$O$_3$ \cite{Keer1976} and 240 J/cm$^3$ for VO$_2$ \cite{Chandrashekhar1973}).
In the remainder of the manuscript we discuss the possible origin of the larger acoustic modulation signal that is observed for V$_2$O$_3$.
First, the electron-phonon coupling coefficient is higher for V$_2$O$_3$ (3x10$^{18}$ W K$^{-1}$m$^{-3}$ ~\cite{supMaterial}) than for VO$_2$ (10$^{18}$ W K$^{-1}$m$^{-3}$ ~\cite{Liu2012}), which partially explains the increased signal strength since energy couples more efficiently between the electrons and the lattice.
This means that the strain wave generation process following electronic photoexcitation is more efficient in V$_2$O$_3$, and that the subsequent strain-induced modifications of the electronic spectral weight are stronger.
In fact, metallic V$_2$O$_3$ at atmospheric conditions is seen to lie near a transition line to a pressure induced isostructural paramagnetic insulating phase (Fig. S1), so that one would indeed expect a similar pressure change to lead to a larger change in the conductivity for V$_2$O$_3$ than for VO$_2$.
Second, the lattice structure of V$_2$O$_3$ is overall more amenable to modulations, as deduced from the lower phonon damping measured in V$_2$O$_3$ compared to VO$_2$ which effectively means that the structure is less rigid.
This implies that photoinduced strain modulations would be larger in V$_2$O$_3$  for the same amount of energy transferred to the lattice.
Lastly, it is important to reiterate the strong influence of differences in defect density, which lead to variations in the static and dynamic properties of both VO$_2$ and V$_2$O$_3$ when comparing nominally equivalent samples.
In particular, such variations are known to affect the structural response of these materials \cite{Ramirez2015}, and should therefore be taken into account when attempting precise quantitative comparisons.

%%% CONCLUSION %%%
\section{CONCLUSION}
\label{conc}

This work demonstrates that strong electron-phonon coupling exists and is responsible for clear ultrafast acoustic modulations of the Drude and optical responses in V$_2$O$_3$ and VO$_2$.
This effect appears to be stronger for V$_2$O$_3$ than for VO$_2$, suggesting that the electronic and lattice structure of V$_2$O$_3$ is more amenable to transient strain modulation than VO$_2$.

We further identify a significant temperature dependence of the sound velocity in both materials, in particular a dramatic increase by a factor of about 4-5 in VO$_2$ and 3-5 in V$_2$O$_3$ across the IMT.
The observation (in both materials) of stronger acoustic damping in the metallic phase relative to the insulating one confirms the strong role of phonon anharmonicity in metallic VO$_2$ and suggests that a similar mechanism is at play in V$_2$O$_3$.
The timescale for damping of the acoustic modulations in both the THz conductivity (Drude response) and the near infrared reflectivity (spectral weight at 1.55 eV, related to the occupation of V3d orbitals) are longer for V$_2$O$_3$ than for VO$_2$, indicating stronger damping in the latter and thus a potentially stronger phonon anharmonicity contribution.

Our findings demonstrate that transient strain induced by photoexcitation is a useful tool to both analyze and control the electronic properties of complex materials and their coupling to lattice excitations.
The same approach could be used to investigate the modifications to the electronic and lattice properties induced by different defect densities.
This is particularly relevant in the case of VO$_2$ and V$_2$O$_3$, where different defect densities lead to strong sample to sample variation of static and dynamic properties, but is likely applicable to other complex materials and to their properties which are characteristically sensitive to small perturbations.

%%% ACKNOWLEDGEMENTS %%%
\section{ACKNOWLEDGEMENTS}

We thank Mariano Trigo and Gabriel Lantz for useful discussions.

E. A. acknowledges support from the ETH Zurich Postdoctoral Fellowship Program and from the Marie Curie Actions for People COFUND Program.
R. D. A. and E. A. acknowledges support from DOE---Basic Energy Sciences under Grant No. DE-FG02-09ER46643.
S. J. Y. and H.-T. K. acknowledge support from the MIT project at ETRI.
I. K. S., S. W., J. G. R. and K. W. acknowledge support from AFOSR under Grant No. FA9550-12-1-0381.
One of us (I.K.S.) acknowledges support from the Vannevar Bush Faculty Fellowship program sponsored by the Basic Research Office of the Assistant Secretary of Defense for Research and Engineering and funded by the Office of Naval Research through grant N00014-15-1-2848.
J. G. R. kindly acknowledges support from FAPA program through Facultad de Ciencias and Vicerrectoria de Investigaciones of Universidad de los Andes, Bogot\'a, Colombia, and Colciencias No. 120471250659.

%%% BIBLIOGRAPHY %%%

\bibliographystyle{apsrev4-1}
\bibliography{VO2V2O3Text}



\section*{Supplementary material}

# I. PHASE DIAGRAMS OF VO$_2$ AND V$_2$O$_3$

Figure S1 shows the phase diagrams of VO$_2$ (Fig. S1a) and V$_2$O$_3$ (Fig. S1b), illustrating the different phases mentioned in the main text[1–6]. The samples studied are undoped and mostly unstrained (except for some minimal interfacial strain from the lattice mismatch with the sapphire substrate, which has a different structure). The effect of heating, triggered by a static temperature increase or via photoexcitation, is indicated by the red dashed line in each diagram. Note in Fig. S1b that as the V$_2$O$_3$ sample temperature increases above 200 K the material approaches a crossover region from a correlated paramagnetic metal to a paramagnetic insulator. This is likely related to the decrease in static conductivity observed in Figure 1b of the main text[7].

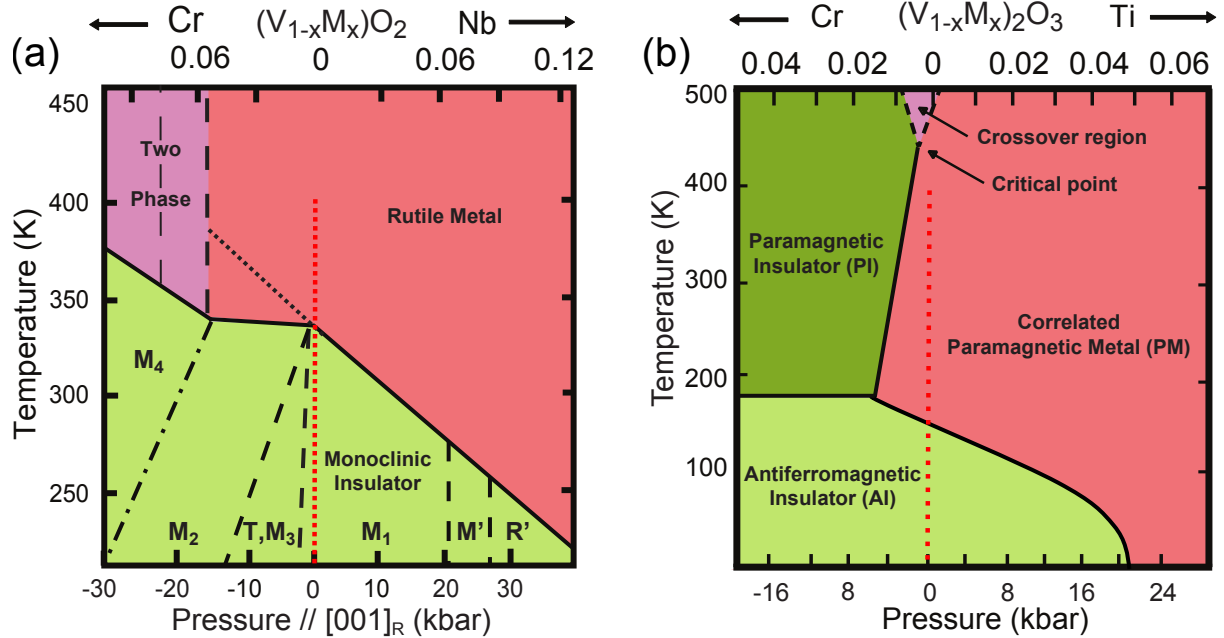

FIG. S1. Phase diagram of single crystalline (a) VO$_2$ and (b) V$_2$O$_3$, adapted from Liu *et al.*[3–6] and Abreu *et al.*[1,2]. Pressure and chemical doping (M = Cr or Nb for VO$_2$, and M = Cr or Ti for V$_2$O$_3$) are approximately equivalent in both compounds[8], as indicated by the corresponding x-axes. The samples are undoped and not purposely strained, so that their properties follow the red dashed lines on the phase diagrams.

The temperature induced IMT in unstrained pure VO$_2$ and pure V$_2$O$_3$ is accompanied by a structural phase transition during which lattice parameters change by about 1%[9,10]. V-O

| | VO$_2$ | | V$_2$O$_3$ | |
|---|---|---|---|---|
| **Parameter** | **I → M [%]** | | **Parameter** | **I → M [%]** |
| Volume | +0.044 | | Volume | -1.4 |
| a$_m$ (c$_R$) | -1.0 | | c$_H$ | +0.2 |
| b$_m$ (a$_R$) | +0.6 | | a$_H$ | -0.9 |
| c$_m$ | -0.2 | | — | — |
| V-O equatorial | -2.1 | | — | — |
| V-O apical | +1.8 | | — | — |
| V-O average | -0.8 | | V-O average | -0.6 |
| V-V average | -3.6 | | V-V average | -2.8 |

TABLE SI. Relative variation of the volume, lattice parameters, V-O bond lengths and V-V bond lengths with increasing temperature across the IM and structural phase transitions, for VO$_2$ (left) and V$_2$O$_3$ (right)[9–12]. VO$_2$ lattice parameters are shown for monoclinic axes (insulating structure), with corresponding rutile axes (metallic structure) where applicable[9]. V$_2$O$_3$ lattice parameters are shown for hexagonal axes, which are true hexagonal axes in the rhombohedral metallic phase and pseudo-hexagonal axes in the monoclinic insulating phase[10].

and V-V bond lengths are also modified by 1-20%[11,12]. Relative changes in these quantities are shown in Table SI, as a reference for the discussion in the main text.

## II. COMPLEMENTARY DATA AND DISCUSSION

### A. Static Far-infrared Conductivity

Figure S2 shows the temperature dependence of the static conductivity for the 95 nm thick V$_2$O$_3$ film. The maximum conductivity of about 1850 $\Omega^{-1}$cm$^{-1}$ is reached at 210 K, following an IMT at $T_{IMT}$ = 175 K. The open black (solid red) symbols show increasing (decreasing) temperature, and the fully metallic region is shaded gray. The single crystallinity and smaller defect density of the 95 nm V$_2$O$_3$ film produces a narrower hysteresis, higher transition temperature and larger metallic conductivity than what is reported in Figure 1b for the polycrystalline 75 nm thick V$_2$O$_3$ film. The conductivity decrease in the metallic phase is

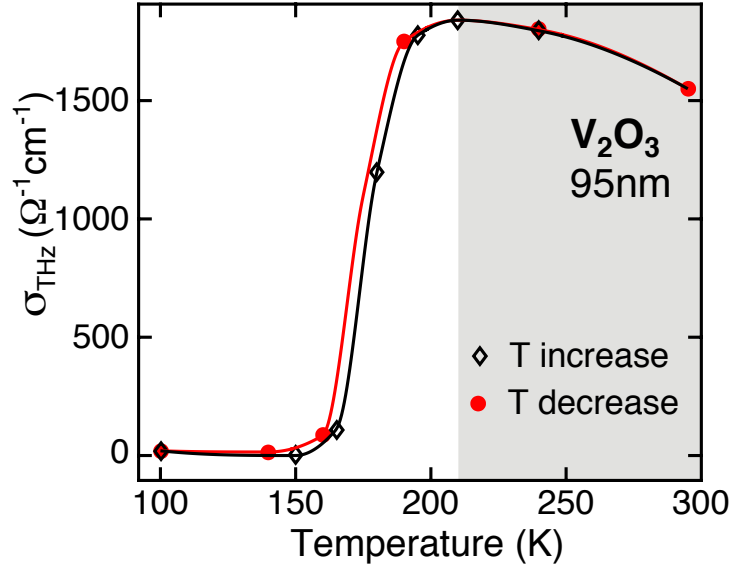

FIG. S2. Temperature dependence of the static far-infrared conductivity for the 95 nm thick $V_2O_3$ thin film, reproduced from Abreu et al.[13]. Open black (solid red) symbols indicate increasing (decreasing) temperature. The sample displays hysteresis, characteristic of the first order phase transition happening at 175 K. The fully metallic region is shaded gray.

comparable between the two samples (10.5 % conductivity drop per 50 K for the 75 nm film vs. 9.2 % conductivity drop per 50 K for the 95 nm film).

Detailed temperature dependent static conductivity measurements are not available for the 50 nm $VO_2$ film, which is however known to have a broader hysteresis than the 75 nm film. The maximum conductivity and $T_{IMT}$ values used in Fig. 4 of the main text were estimated by comparing the low temperature $\Delta\sigma$ behavior for the 50 nm (Fig. S6b) and 75 nm (Figs. 2a and S6a) films. The maximum $\Delta\sigma$ for the 50 nm film (1100 $\Omega^{-1}$cm$^{-1}$) is about half that for the 75 nm (2400 $\Omega^{-1}$cm$^{-1}$). Also, $\Delta\sigma$ decreases to ∼20% of its maximum value, due to the transition into the metallic phase, at a temperature of about 350 K for the 75 nm film and 330 K for the 50 nm film, i.e. 20 K lower for the latter. By comparing these values to the static data for the 75 nm thick film (Fig. 1a of the main text), a maximum conductivity of $\frac{4800}{2} = 2400$ $\Omega^{-1}$cm$^{-1}$ and a $T_{IMT} = 360 - 20 = 340$ K can therefore be estimated for the 50 nm film. Given the broader hysteresis of the 50 nm film $T_{IMT}$ could be overestimated by up to 30 K, which would have no significant impact of the conclusions drawn from Fig. 4c of the main text. In general, variations in transition temperatures with

4

film thickness is not uncommon in vanadates and can have several origins, the most common being slightly different strain environments[14,15].

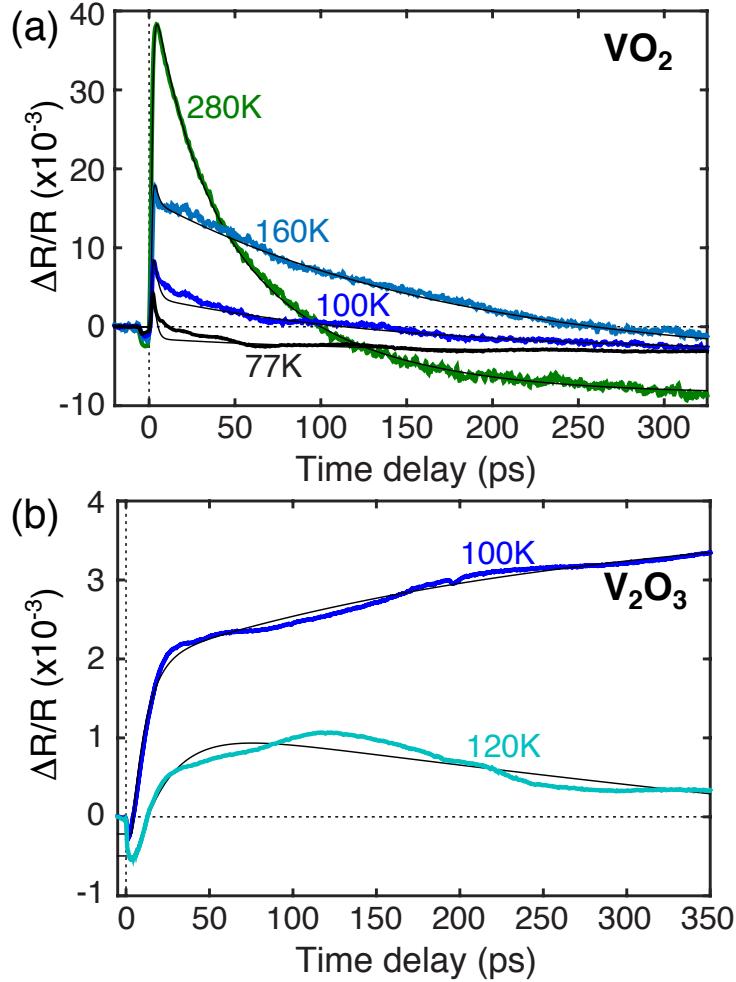

FIG. S3. Transient $\Delta R/R$ in the 75 nm (a) $VO_2$ and (b) $V_2O_3$ films for temperatures below $T_{IMT}$ at pump fluences of 3.8 mJ/cm$^2$ and 1 mJ/cm$^2$, respectively. Superimposed onto the data are black lines corresponding to fits to a sum of exponentials, as detailed in the text. The background-subtracted acoustic $\Delta R/R$ oscillations derived from this data are shown in Figs. 5a and 5d of the main text.

### B. Low temperature $VO_2$ and $V_2O_3$ reflectivity dynamics

The full set of low temperature $\Delta R/R$ data used to extract the acoustic responses shown in Figs. 5a and 5d of the main text is depicted in Fig. S3. The overall optical reflectivity

5

dynamics are quite complex, especially close to the IMT. An initial sub-picosecond transient in $\Delta R/R$, due to carrier photoexcitation across the insulating bandgap and subsequent electronic thermalization, is observed in both materials. This is followed by a transient response with a characteristic timescale on the picosecond scale, due to equilibration of the electronic temperature with that of the lattice. Subsequent $\Delta R/R$ dynamics track the evolution of the 1.55 eV spectral weight, due essentially to V3d - V3d transitions, as an increasingly large metallic volume fraction is created following photoexcitation (leading to an increasing conductivity, as seen in Fig. S6). The slower dynamics of the metallic phase are the most sensitive to the acoustic modulation effects discussed here.

The exact details of the $\Delta R/R$ dynamics near the IMT goes beyond the scope of this work[16]. For the purpose of our analysis, the background dynamics for both materials can be phenomenologically fit by a sum of at most three exponentials. These fits are shown as black lines in Fig. S3. The slow acoustic oscillations are isolated by subtracting this exponentially decaying background from the raw data. The resulting acoustic oscillations are shown in Figs. 5a and 5d of the main text.

### C. High temperature VO$_2$ and V$_2$O$_3$ reflectivity and conductivity dynamics

Figure S4 depicts the full time window for the $T > T_{IMT}$ data shown in Figs. 2-5 of the main text, for all four films (50 nm and 75 nm thick VO$_2$, and 75 nm and 95 nm thick V$_2$O$_3$). Figures S4a and S4b show the metallic $\Delta R/R$ and $\Delta\sigma$ transient response for the 75 nm thick VO$_2$ film, respectively. Figure S4c shows the $\Delta\sigma$ transient response for the 50 nm thick VO$_2$ film. Figures S4d and S4e show the metallic $\Delta R/R$ and $\Delta\sigma$ transient response for the 75 nm thick V$_2$O$_3$ film, respectively. Figure S4f shows the $\Delta\sigma$ transient response for the 95 nm thick V$_2$O$_3$ film. The black lines are the fits used to extract the acoustic response as discussed in Section II B.

### D. Fitting $\Delta R/R$ with a damped sinusoid, damping timescale

After subtracting the exponential dynamics described above, and shown as black lines in Figs. S3 and S4, one obtains the oscillatory behavior depicted in Figs. 4 and 5 of the main text. For the 75 nm films, the $\Delta$R/R oscillatory data from Figs. 5a, 5b, 5d and 5e are fit to

6

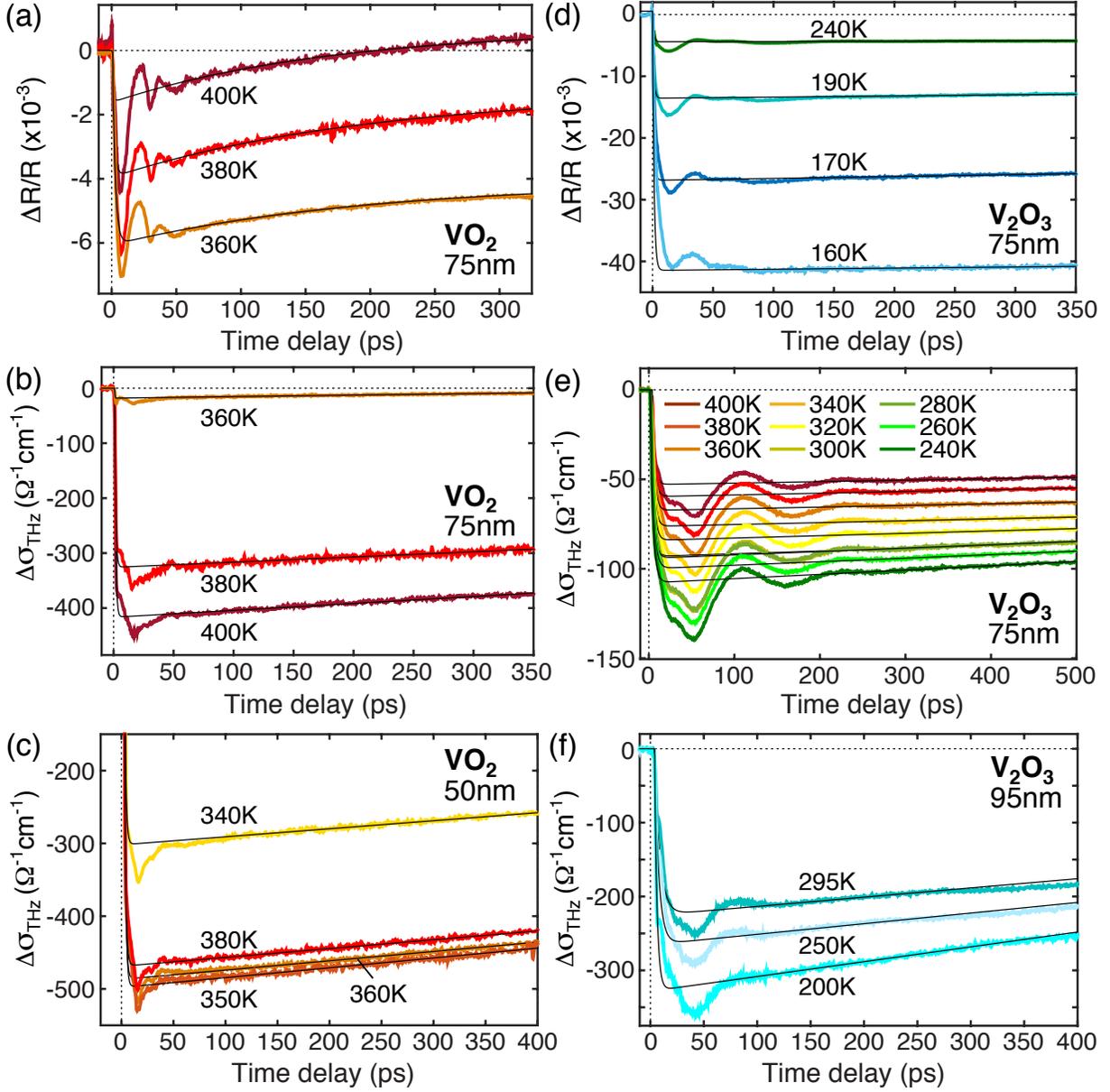

FIG. S4. Full length transient $\Delta R/R$ (a) and $\Delta\sigma$ (b and c) of VO$_2$, and $\Delta R/R$ (d) and $\Delta\sigma$ (e and f) of V$_2$O$_3$ for temperatures above $T_{IMT}$. The results correspond to a 75 nm film (a and b) and to a 50 nm film (c), both at a pump fluence of 3.8 mJ/cm$^2$. (d) and (e) correspond to a 75 nm film at a pump fluence of 1 mJ/cm$^2$, while (f) corresponds to a 95 nm film at a pump fluence of 3 mJ/cm$^2$. Superimposed onto the data are black lines corresponding to fits to a sum of exponentials, as detailed in the text.

extract both the oscillation periods (Figs. 5c and 5f) and the damping times (Fig. S5). The chosen fitting function is a damped sinusoid, $R = R_0 + A \times \exp(-t/t_0) \times \sin(2\pi(t-t_c)/\tau)$,

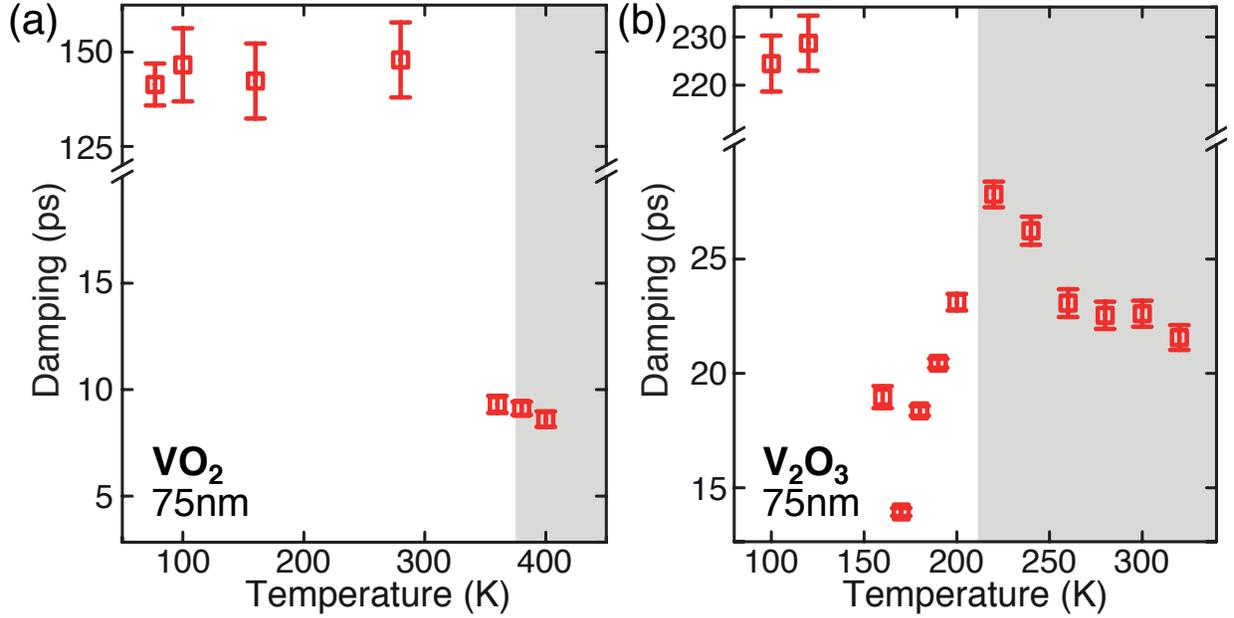

FIG. S5. Damping times extracted from fitting the acoustic component of ΔR/R dynamics data (shown in Figs. 5a, 5b, 5d and 5e of the main text) to a damped sinusoid. Data are shown for 75 nm thick films of (a) $VO_2$ and (b) $V_2O_3$. The fully metallic phase is shaded gray, for comparison with Fig. 1 of the main text.

where $\tau$ is the period and $t_o$ is the damping time. Table SII lists the adjusted $R^2$ parameters obtained for the fits. These are mostly very close to 1, attesting to the reliability of the damped sinusoid model in describing our data. $VO_2$ data in the 100 - 160 K range has slightly lower but still very acceptable $R^2$ parameters due to a decreased signal-to-noise ratio.

Figure S5 shows the temperature dependent damping times (equivalent to the inverse of the scattering rate) of $VO_2$ (Fig. S5a) and $V_2O_3$ (Fig. S5b). Interestingly, we measure a much slower acoustic damping time in the insulating phase of $VO_2$ compared to the metallic phase (cf. Figs. 5a and 5b of the main text), consistent with the work of Budai et al.[17] which shows that phonon-phonon scattering present in the metallic phase essentially disappears in the insulating phase of $VO_2$. The same temperature dependence of the damping is observed for $V_2O_3$, with lower values in the metallic phase (Figs. 5d and 5e). An analysis similar to that of Budai et al.[17], done on $V_2O_3$, would provide a measure of the importance of phonon entropy effects in the metallic phase of this material, as compared to $VO_2$.

It is interesting to note that the damping time for $V_2O_3$ increases between about 150 K

| V$_2$O$_3$ | | VO$_2$ | |
|---|---|---|---|
| T (K) | R$^2$ (%) | T (K) | R$^2$ (%) |
| 100 | 94.8 | 77 | 88.4 |
| 120 | 97.5 | 100 | 71.7 |
| 160 | 99.1 | 160 | 68.9 |
| 170 | 99.8 | 280 | 68.3 |
| 180 | 99.8 | 360 | 99.5 |
| 190 | 99.9 | 380 | 99.7 |
| 200 | 99.7 | 400 | 99.1 |
| 220 | 99.7 | | |
| 240 | 99.6 | | |
| 260 | 99.4 | | |
| 280 | 99.3 | | |
| 300 | 99.5 | | |
| 320 | 99.4 | | |

TABLE SII. Adjusted R$^2$ parameters obtained for the fits of the acoustic component of $\Delta R/R$ dynamics data with a damped sinusoid, for V$_2$O$_3$ (left) and VO$_2$ (right).

and the start of the shaded region, at 220 K, where full metallic conductivity is reached. Insulating and metallic volume fractions coexist in this 150 - 220 K temperature interval (cf. Fig. 1b in the main text). Insulating regions hinder the propagation of acoustic modulations throughout the metallic volume, thereby decreasing the damping time in the (lower) temperature region where the insulating volume fraction is larger[18]. An increasing damping time in the coexistence region could also be a signature of an intermediate monoclinic metallic phase[19–21]. As the system becomes fully metallic (shaded area), and while it remains in the ∼50 K wide (lower) temperature region where the conductivity is maximal and approximately constant (Fig. 1b), the damping becomes slightly faster as temperature increases — potentially the result of phonon-phonon interactions increasing with temperature. Above 250 K the damping time remains essentially constant until 320 K, the limit of our measurement. A temperature independent damping in the fully metallic phase of V$_2$O$_3$ is also consistent with the results of Budai *et al.* on VO$_2$[17], where this effect was

interpreted as a signature of electron-phonon effects contributing to phonon anharmonicity in addition to simply phonon-phonon interactions. Our data also yields a stable damping time for metallic $VO_2$, over the small temperature interval available from our measurements. The temperature independence of acoustic damping properties in the metallic phase of both materials therefore lends further support to the main message of this manuscript, namely the significance of electron-phonon coupling in $VO_2$ and $V_2O_3$.

As shown explicitly in Fig. S5, the acoustic scattering rate is higher in both phases of $VO_2$ compared to $V_2O_3$. This can also be seen directly from Figs. 4 and 5 of the main text, where the acoustic modulations persist longer for $V_2O_3$ than for $VO_2$. Indeed, the conductivity dynamics (Fig. 4) show that the oscillations essentially disappear 300 ps after photoexcitation for $V_2O_3$, whereas they have already disappeared around 100 ps in the case of $VO_2$. Similarly, the reflectivity dynamics (Fig. 5) show a strong decrease in the oscillation amplitude at about 150 ps for $V_2O_3$, and at about 75 ps for $VO_2$.

The sensitivity of conductivity and reflectivity dynamics measurements differs but the conclusion is the same - there are fewer decay channels for acoustic excitation in $V_2O_3$ compared to $VO_2$. Budai *et al.*[17] have reported that the fast phonon damping in $VO_2$ arises from a large anharmonicity in the lattice potential. Our present observation suggests a lower degree of anharmonicity for the lattice potential in $V_2O_3$, compared to $VO_2$.

### E. Low temperature $VO_2$ and $V_2O_3$ conductivity dynamics

Figure S6 shows the full time window of the temperature dependent $\Delta\sigma(t)$ for $T < T_{IMT}$, for all four films discussed in the main paper: 75 nm $VO_2$ (Fig. S6a), 50 nm $VO_2$ (Fig. S6b), 75 nm $V_2O_3$ (Fig. S6c) and 95 nm $V_2O_3$ (Fig. S6d). First, it is clear that for initial temperatures well below $T_{IMT}$ lower initial temperatures lead to lower photoinduced conductivities, regardless of the film thickness or the specific vanadium oxide. For initial temperatures approaching $T_{IMT}$ the film is already partly metallic and the transient conductivity starts to decrease. Second, the onset of the conductivity recovery is clearly seen for the lowest temperatures within the time window, indicating that only a metastable and partially metallic state is achieved. These behaviors are observed in all but Fig. S6c, where the minimum investigated temperature is already quite close to $T_{IMT}$.

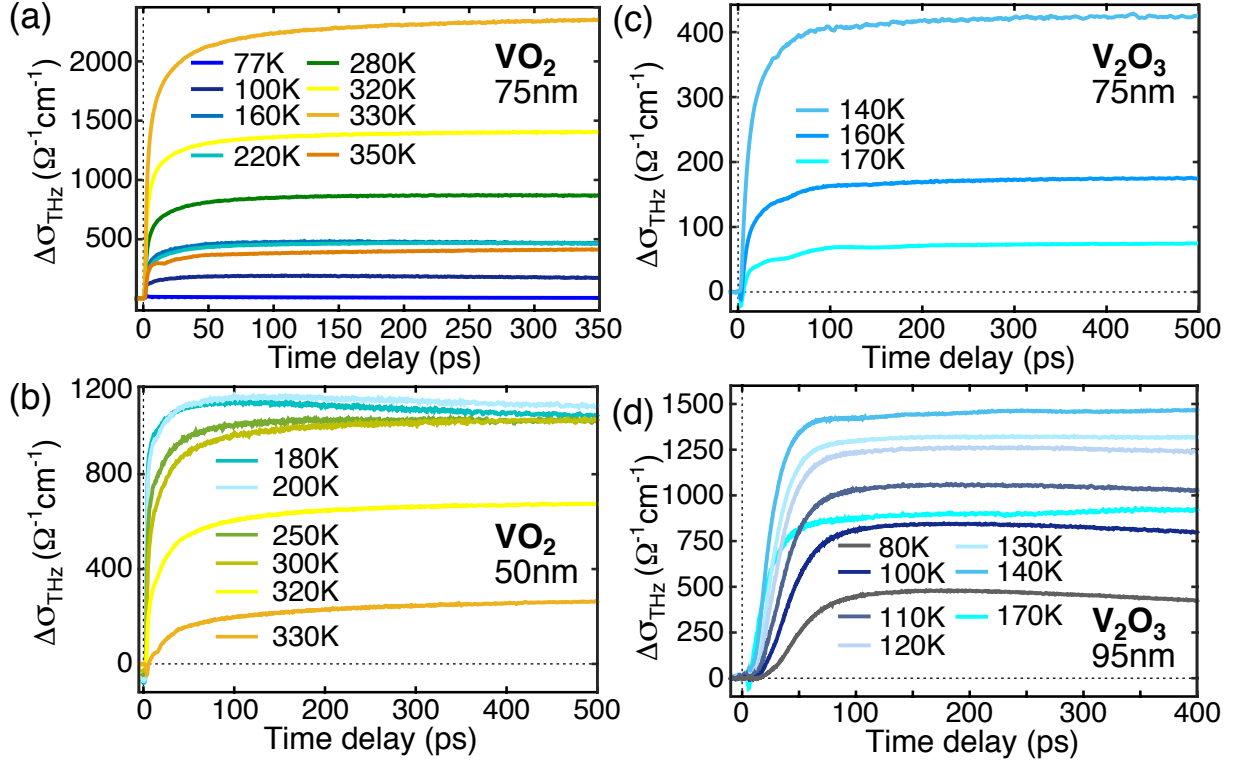

FIG. S6. Conductivity dynamics at $T < T_{IMT}$ for the (a) 75 nm $VO_2$, (b) 50 nm $VO_2$, (c) 75 nm $V_2O_3$, and (d) 95 nm $V_2O_3$ films. Pump fluences were 3.8 mJ/cm$^2$ in (a) and (b), 1 mJ/cm$^2$ in (c), and 3 mJ/cm$^2$ in (d). The final conductivity as well as the lifetime of the photoexcited metallic state increases with initial temperature. For temperatures close to $T_{IMT}$ the transient conductivity starts to decrease since the film is already partly metallic.

### F. Comparison of different film thicknesses

Figure S7 shows the comparison between different film thicknesses for a given material. In Fig. S7a, the acoustic component of $\Delta\sigma(t)$ for the two $VO_2$ films with different thicknesses of the same orientation and similar characteristics are compared. The acoustic oscillations are clearly faster in the 50 nm film, consistent with a shorter time needed for the acoustic perturbation to cross a thinner film at a comparable sound velocity. The same conclusion can be drawn even more directly from the $\Delta R/R$ dynamics shown in Fig. S7b where a normalized plot comparing the acoustic oscillations for both the 50 nm and 75 nm thicknesses at 360 K is shown. This confirms the validity of our analysis where we relate the oscillation period to the sound velocity in the system.

11

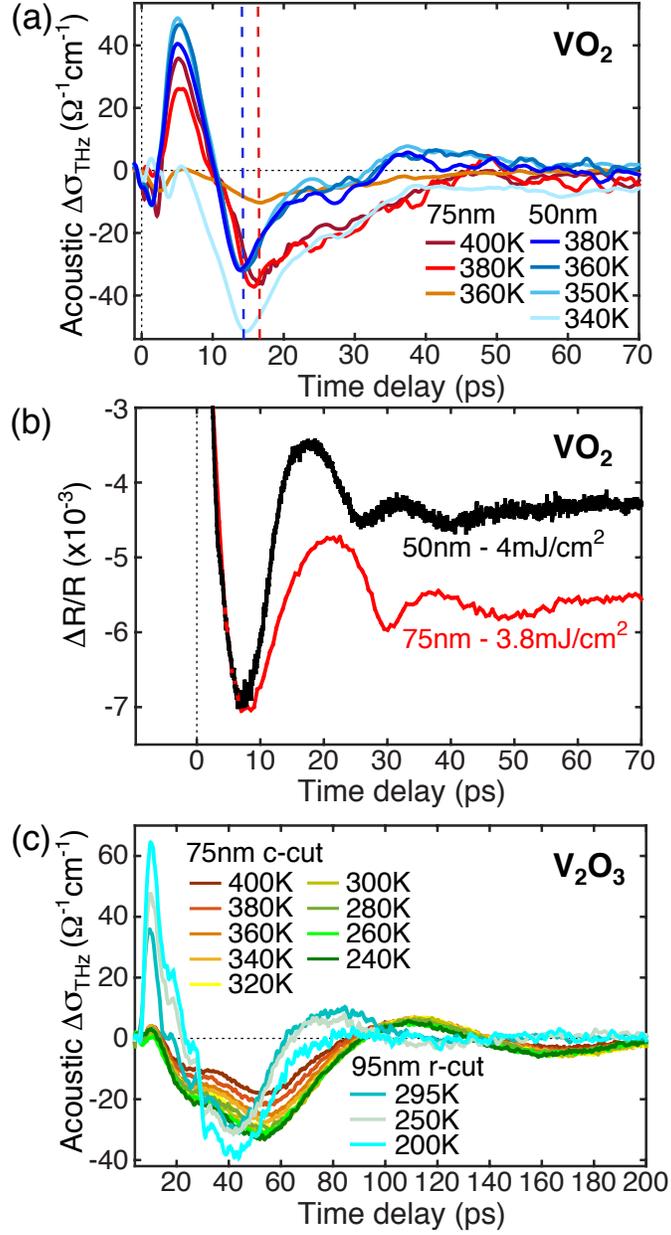

FIG. S7. Comparison for different film thicknesses. Background subtracted $\Delta\sigma$ acoustic oscillations for (a) VO$_2$ and (c) V$_2$O$_3$, corresponding to the data in (a) Fig. S4b and S4c, and (c) Figs. S4e and S4f. Two film thicknesses are compared for both materials, 75 nm and 50 nm for VO$_2$, and 75 nm and 95 nm for V$_2$O$_3$, which exhibit different delays between acoustic oscillations, as discussed in the text. (b) shows $\Delta R/R$ oscillations at 360 K for the same VO$_2$ films, confirming the shorter delay between oscillations in the thinner film. $\Delta\sigma(t)$ data for both VO$_2$ films (a) and for the 95 nm V$_2$O$_3$ film (c) were smoothed using a five point (1 ps) moving average.

Figure S7c shows the acoustic component of $\Delta\sigma$ for the two $V_2O_3$ films. In this case it is difficult to compare the oscillation period since the films have different orientations and growth conditions, leading to different properties. In fact, the reflected acoustic signal is faster for the thicker film, contrary to what would be expected and to what is observed in the case of $VO_2$. This discrepancy in the $V_2O_3$ films can be explained by a combination of 1) a different sound propagation direction due to different crystal cuts, 2) different substrate-lattice mismatch in both films, and 3) different degrees of crystallinity. All of these lead to variations in the sound velocity normal to the sample, so that the oscillation period in the two $V_2O_3$ films cannot be directly compared.

A detailed description beyond the present qualitative discussion of the many additional contributions to the photoinduced acoustic response in such thin films (with thickness on the order of the acoustic wavelength) is outside the scope of the present work.

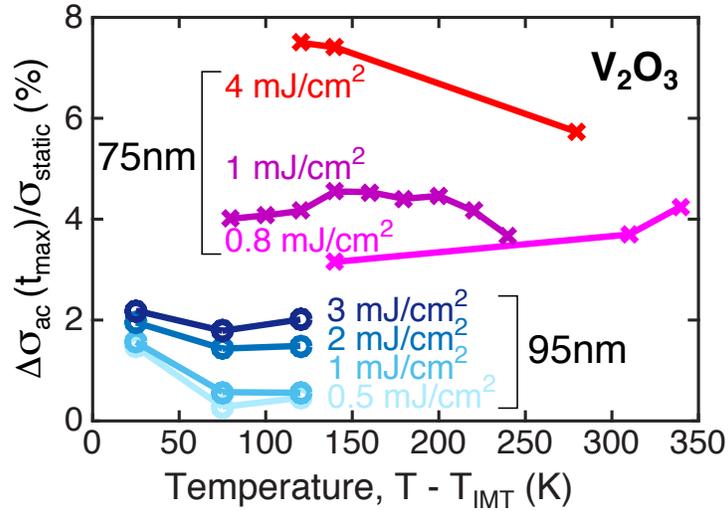

FIG. S8. Fluence dependence of the $V_2O_3$ conductivity modulation data shown in Fig. 4 of the main text. Data are shown for the 75 nm $V_2O_3$ film (red/pink) for fluences in the 0.8 - 4 mJ/cm$^2$ range, for the 95 nm $V_2O_3$ film (blue) for fluences in the 0.5 - 3 mJ/cm$^2$ range. As expected, the acoustic response increases with photoexcitation fluence.

### G. Fluence dependence

Figure S8 depicts the fluence dependence of the $V_2O_3$ conductivity data shown in Fig. 4 of the main text. As expected, the induced acoustic response increases with fluence. Since

13

the latent heat of $V_2O_3$ is significantly lower than that of $VO_2$, lower fluences were used for the $V_2O_3$ measurements at low temperatures. This was maintained for the high temperature measurements, for consistency. Also, the damage threshold for $V_2O_3$ is five times lower than that for $VO_2$[3,22], which limits the fluence values that can be used on $V_2O_3$ in general. To more effectively compare the observed acoustic modulation of the conductivity between the two materials in Fig. 4 of the main text, 3 mJ/cm$^2$ $V_2O_3$ data on the 95 nm film were chosen, being the closest match to the 3.8 mJ/cm$^2$ used for the $VO_2$ 75 nm film. 1 mJ/cm$^2$ data were chosen for the 75 nm $V_2O_3$ film since this is the fluence at which more data points were available for this sample. While it is not possible to deduce a precise functional fluence dependence of the acoustic modulation from this data, the results in Fig. S8 show that the modulation induced in $V_2O_3$ with a pump fluence of 4 mJ/cm$^2$ is about 1.7 times higher than that induced with 1 mJ/cm$^2$. Figure 4 in the main text shows that the acoustic modulation in 75 nm thick $V_2O_3$ following 1 mJ/cm$^2$ photoexcitation is larger than that in 75 nm thick $VO_2$ following 3.8 mJ/cm$^2$ photoexcitation. The modulation in 75 nm thick $V_2O_3$ following 4 mJ/cm$^2$ photoexcitation would therefore be even larger. Consequently, the conclusion that the acoustic modulation in $V_2O_3$ is stronger than that in $VO_2$ (in the main text, based on Fig. 4) is not affected by the fact that the analysis is done at different fluences.

To be precise in comparing pump induced effects one should consider deposited energy density in the material rather than only incident fluence. In the case of metallic $VO_2$ and $V_2O_3$, however, the penetration depth and reflectivity of the 800 nm wavelength pump beam are very similar, as seen in Table SIII. The slight reduction in absorbed fluence due to a higher reflectivity in $VO_2$ is compensated by a smaller penetration depth, so that overall fluence values can be directly compared in the metallic phases of the two materials. As a side note, comparing incident fluences at 800 nm is still equivalent to comparing deposited energy densities even in the insulating phase of $VO_2$ and $V_2O_3$ (Table SIII).

## III. DETAILED ANALYSIS OF A THICKER FILM

We also performed reflectivity measurements on a thicker sample, a 400 nm $V_2O_3$ film deposited on 70 nm of sapphire with a bulk silicon substrate[25], for comparison. Figure S9 shows the corresponding $\Delta R/R$ dynamics at three temperatures above $T_{IMT}$, for a fluence

|  | Material | Penetration depth | Reflectivity |
|---|---|---|---|
| Metallic | $VO_2$ | 74 nm | 0.19 |
|  | $V_2O_3$ | 99 nm | 0.12 |
| Insulating | $VO_2$ | 82 nm | 0.26 |
|  | $V_2O_3$ | 171 nm | 0.13 |

TABLE SIII. Penetration depth and reflectivity values for $VO_2$ and $V_2O_3$, estimated from dielectric constant measurements[23,24].

of 1 mJ/cm². It is clear from this figure that for this larger film thickness, the dynamics are dominated by acoustic echoes of the strain pulse reflecting off the interface. Here a model of the propagating acoustic discontinuity can be attempted, as briefly hinted at in the supporting information of Liu *et al.*[3].

In detail, we observe several dynamical stages in Fig. S9. The sub-picosecond peak in (a) is due to electron-electron collisions (ballistic transport) and electronic thermalization. The non-uniform heat absorption across the film generates a coherent strain wave ($L_1$) which propagates along the film and then reflects from the sample-substrate interface ($L_2$, at 100 ∼ 150 ps). Subsequent acoustic reflections after ∼ 150 ps are too small to be detected.

Figure S9b shows a fit of the 220 K $\Delta R/R$ data using the two-temperature model (TTM)[26] which tracks the electronic ($T_e$) and lattice ($T_l$) temperatures in the first 1.5 ps after excitation,

$$C_e(T_e)\frac{\partial T_e}{\partial t} = -g(T_e - T_l) + P(t)$$
$$C_l\frac{\partial T_l}{\partial t} = g(T_e - T_l).$$

Here, the electron specific heat is given by $C_e(T_e) = \gamma T_e$, where $\gamma$ = 80 mJ K$^{-2}$mol$^{-1}$, and the lattice specific heat is taken as constant, $C_l$ = 100 J K$^{-1}$mol$^{-1}$. P(t) denotes the absorbed pump power density, which accounts for the absorption length of the pump pulse in the sample (cf. discussion of the penetration depth in Section II G, above). These assumptions are valid above $T_{IMT}$, i.e. when the latent heat contribution is no longer relevant. The only free parameter is the $V_2O_3$ electron-phonon coupling constant g, for which the TTM fit yields a large value of g = 3×10$^{18}$ W K$^{-1}$m$^{-3}$.

15

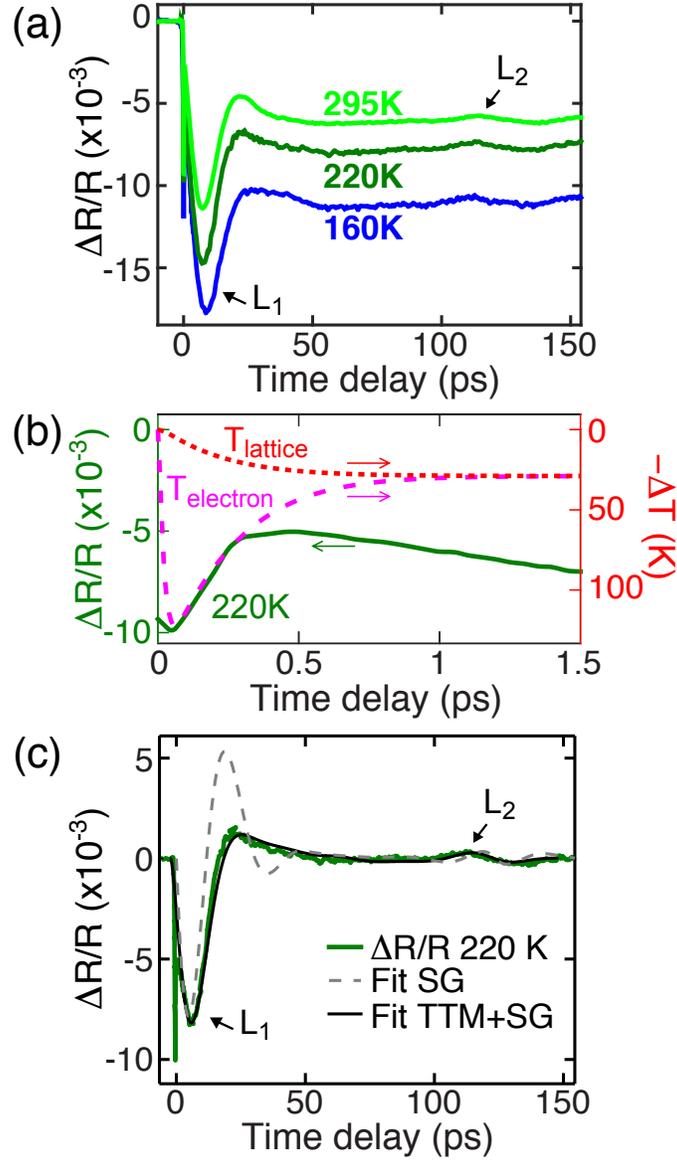

FIG. S9. (a) Reflectivity change $\Delta R/R$ in the 400 nm $V_2O_3$ film, showing a fast electronic response followed by two acoustic waves signatures, $L_1$ and $L_2$, due to the main acoustic wave and to its reflection from the sample-substrate interface. (b) Two-temperature model fit of the 220 K data (green line, left-hand side y-axis) between 0 and 1.5 ps, yielding an electron-phonon coupling constant of $g = 3\times10^{18}$ W $K^{-1}m^{-3}$. Electronic and lattice temperatures are shown as dashed pink and dotted red lines, respectively, and correspond to the right-hand side y-axis. (c) Fit of the 220 K data (green line) with the strain generation only (SG) model (dashed gray lines) and with the combined two-temperature with strain generation (TTM+SG) model (full black lines).

As shown in Fig. S9b, the TTM successfully fits the electronic response in the initial ∼0.5 ps, where the electronic temperature constitutes the main contribution to the dynamics of the system. Fitting in this region is sufficient to determine g. The lattice simultaneously heats, as illustrated by the evolution of $T_{lattice}$ in Fig. S9b. Were the response of the lattice purely thermal, as assumed by the TTM, the reflectivity dynamics would track the dynamics of $T_{lattice}$ (and $T_{electron}$). However, some of the energy is transferred to different modes such as the acoustic ones which constitute the focus of this paper. These modulate the spectral weight at 1.55 eV, and consequently the reflectivity dynamics, such that the $\Delta R/R$ diverges from a simple exponentially decaying thermal response, as clearly seen in the larger time window shown in Fig. S9a.

In Figure S9c, the acoustic wave propagation during the first 150 ps is simulated using two different models. The first is a relatively simple strain generation (SG) model[27,28], given by

$$\rho \frac{\partial^2 u}{\partial t^2} = 3\frac{1-\nu}{1+\nu}B\frac{\partial^2 u}{\partial z^2} - 3B\beta\frac{\partial \nabla T(z)}{\partial z},$$

where $\rho = 5 \times 10^3$ kg/m$^3$ is the density of the V$_2$O$_3$ film, $u$ is the displacement of the lattice normal to the film surface, $\nu = 0.33$ is the Poisson ratio[29], B $= 2 \times 10^{11}$ Pa is the bulk modulus of V$_2$O$_3$[30], z is the acoustic wave propagation direction, $\beta = 2.5 \times 10^{-6}$ is the linear expansion coefficient[31], and $\nabla$T is the time-independent exponentially decaying temperature gradient across the film, which accounts for the absorption length of the pump pulse.

The displacement $u(z,t)$ obtained from this expression is used to estimate the transient reflectivity amplitude $\Delta R/R$ from

$$\eta = \frac{\partial u}{\partial z}$$
$$\Delta R(t) = \int_0^\infty f(z)\eta(z,t)dz.$$

f(z) is the sensitivity function defined by Thomsen *et al.*[28], in which there are two fitting parameters, $\frac{\partial n}{\partial \eta}$ and $\frac{\partial k}{\partial \eta}$. $n$ and $k$ stand for the real and imaginary parts of the refractive index at 800 nm, given by n = 1.69 and k = 0.59 at 220 K[23]. The sound velocity of the

17

longitudinal strain wave in the film is taken as 8x10³ m/s[32,33]. An additional dimensionless fitting parameter is added to account for the attenuation of the acoustic wave upon reflection at the film-substrate interface.

As seen in Fig. S9c (dashed gray lines), this model provides reasonable fits to the data within the first few ps after photoexcitation, but the predicted wave amplitude at about 25 ps is too large to accurately describe the results. The discrepancy in the simulated response is the result of the model not accounting for the non-equilibrium processes at short time delays nor for electron-lattice thermalization in the thin film.

To correct the discrepancy of the first model, Fig. S9c plots also the prediction from a second model, which combines the two-temperature and the strain generation analyses (TTM+SG):

$$C_e(T_e)\frac{\partial T_e}{\partial t} = \nabla(\kappa_e \nabla T_e) - g(T_e - T_l) + P(z,t)$$
$$C_l\frac{\partial T_l}{\partial t} = \nabla(\kappa_l \nabla T_l) + g(T_e - T_l)$$
$$\rho\frac{\partial^2 u}{\partial t^2} = 3\frac{1-\nu}{1+\nu}B\frac{\partial^2 u}{\partial z^2} - 3B\beta\frac{\partial \nabla T(z,t)}{\partial z},$$

where $\kappa_e = 300$ kg m⁻¹ K⁻¹ and $\kappa_l = 30$ kg m⁻¹ K⁻¹ are the thermal conductivities of the electrons and the lattice, respectively.

This model, which accounts for thermalization effects, is significantly more successful in describing our data (dotted gray lines in Fig. S9c), given the choice of $\frac{\partial n}{\partial \eta}$=-1.63$\frac{\partial k}{\partial \eta}$, than the individual TTM and SG models.

---

[*] elsabreu@phys.ethz.ch. Previously at Department of Physics, Boston University, Boston MA 02215, USA

[†] Previously at Department of Physics, Boston University, Boston MA 02215, USA

[‡] mengkun.liu@stonybrook.edu. Previously at Department of Physics, Boston University, Boston MA 02215, USA

[§] raveritt@physics.ucsd.edu. Previously at Department of Physics, Boston University, Boston MA 02215, USA



\end{document}